\documentclass[runningheads,a4paper]{llncs}
\usepackage{algorithm,algorithmicx,algpseudocode}
\usepackage{amssymb}
\usepackage{graphicx}
\usepackage{color}
\usepackage{amsmath}
\usepackage{multirow}
\usepackage{url}
\usepackage[caption=false]{subfig}
\algrenewcommand{\algorithmiccomment}[1]{\hfill$/*$ #1}

\urldef{\mailsa}\path|{alfred.hofmann, ursula.barth, ingrid.haas, frank.holzwarth,|
\urldef{\mailsb}\path|anna.kramer, leonie.kunz, christine.reiss, nicole.sator,|
\urldef{\mailsc}\path|erika.siebert-cole, peter.strasser, lncs}@springer.com|

\begin{document}

\mainmatter  

\title{Reciprocal Recommendation System for Online Dating}


%
%
\author{Peng Xia\inst{1} \and Benyuan Liu\inst{1} \and Yizhou Sun\inst{2} \and Cindy Chen\inst{1}}

%

\institute{Department of Computer Science, University of Massachusetts Lowell, Massachusetts, USA\\ \and
College of Computer and Information Science, Northeastern University
}

%
%

\toctitle{Lecture Notes in Computer Science}
\tocauthor{Authors' Instructions}
\maketitle

\begin{abstract}
Online dating sites have become popular platforms for people to look for potential romantic partners. Different from traditional user-item recommendations where the goal is to match items (e.g., books, videos, etc) with a user's interests, a recommendation system for online dating aims to match people who are mutually interested in and likely to communicate with each other. We introduce similarity measures that capture the unique features and characteristics of the online dating network, for example, the interest similarity between two users if they send messages to same users, and attractiveness similarity if they receive messages from same users. A reciprocal score that measures the compatibility between a user and each potential dating candidate is computed and the recommendation list is generated to include users with top scores. The performance of our proposed recommendation system is evaluated on a real-world dataset from a major online dating site in China. The results show that our recommendation algorithms significantly outperform previously proposed approaches, and the collaborative filtering-based algorithms achieve much better performance than content-based algorithms in both precision and recall. Our results also reveal interesting behavioral difference between male and female users when it comes to looking for potential dates. In particular, males tend to be focused on their own interest and oblivious towards their attractiveness to potential dates, while females are more conscientious to their own attractiveness to the other side of the line.

\end{abstract}

\section{Introduction}
Online dating sites have become popular platforms for people to look for potential romantic partners, offering an unprecedented level of access to possible dates that is otherwise not available through traditional means. According to a recent survey\footnote{http://statisticbrain.com/online-dating-statistics}, 
40 million single people (out of 54 million) in the US have signed up with various online dating sites such as Match.com, eHarmony, etc, and around 20\%  of currently committed romantic relationships began online, which is more than through any means other than meeting through friends.

Many online dating sites provide suggestions on compatible partners based on their proprietary matching algorithms. Unlike in traditional user-item recommendation systems where the goal is typically to predict a user's opinion towards passive items (e.g., books, movies, etc), when making recommendation of potential dates to a user (referred to as service user) on an online dating site, it is important that not only the recommended users match the user's dating interest, but also the recommended users are interested in the service user and thus likely to reciprocate when contacted. A successful online dating recommendation system should match users with mutual interest in each other and hence result in better chances of interactions between them and improved user satisfaction level. 

In this paper, we study the reciprocal online dating recommendation system based on a large real-world dataset obtained from a major online dating site in China with a total number of 60 million registered users. In particular, \textit{given a user, we seek to identify a set of users who are most likely to be contacted by the service user when recommended and reciprocate when contacted.} 

The characteristics of the our online dating network present unique opportunities and challenges to the reciprocal recommendation problem. First, there is a rich set of user attributes available in our dataset that can be used in the prediction models. These include a user's age, gender, height, weight, education level, income level, house ownership, geographic location, occupation, interests/hobbies, number of photos, etc. In addition, there are a variety of online dating specific information including a user's preference in potential dates (age range, height range, education level, income range, geography location, etc), and his/her dating and marriage plan (when to get married, whether to live with parents and have child after marriage, marriage ceremony style, etc). Moreover, our dataset contains the communication traces between users, i.e., who sent or replied to messages to whom and the associated timestamps.  As shown in our earlier paper \cite{Peng2014}, the communication trace of a user reflects his/her actual dating preference, which may significantly deviate from his/her stated preference and thus should play an important role in the design of the recommendation system. 

Due to the heterogeneous dating nature in our recommendation problem (dating is restricted to users of opposite genders on the online dating site in our study), previous approaches designed for friend recommendation in conventional online social networks such as Facebook and LinkedIn cannot be directly applied. For example, the number of common neighbors is often used for friend recommendation for conventional social networks, i.e., the more common friends two users share, the more likely they will become friends and thus should be recommended to each other. On a heterosexual online dating site, however, a user is only interested in contacting other users of opposite gender, resulting in a bipartite network between males and females. There is no common neighbors between a service user and recommended users since they are of different genders. To this end, we will need to devise appropriate mechanisms that accounts for the special characteristics of the online dating network.

The contributions of this paper are summarized as follows. 

\begin{itemize}
	
\item We present a recommendation system that aims to match users who are most likely to communicate with each other. We introduce similarity measures that capture the unique features and characteristics of the heterogeneous online dating network. In particular, we build a preference model for each service user based on the attributes of users who have been contacted by the service user. We also characterize the interest similarity between two users if they send messages to same users, and attractiveness similarity if they receive messages from same users. Based on these similarity measures we compute the compatibility between a service user and potential dating candidates and the recommendation list is generated to include candidates with top scores. Using a combination of similarity measures, we construct a set of content-based and collaborative filtering-based algorithms with different measures of compatibility between users.
 
\item We evaluate the performance of our proposed algorithms on the real-world online dating dataset that consists of 200,000 users and around two million messages over a period of two months. Our results show that both content-based and collaborative filtering-based recommendation algorithms presented in our paper significantly outperform previously proposed approaches. Also, compared to the content-based algorithms, our collaborative filtering-based algorithms achieve much better performance in both precision and recall.  Moreover, most of our proposed recommendation algorithms place the relevant recommendations (users who have been actually contacted by and replied to the service user) in the top 30\% to 50\% positions of the recommendation list. This is an important performance measure as users tend to look at the list from top to bottom.

\item Our results also reveal interesting behavioral difference between male and female users when it comes to looking for potential dates. Among the collaborative filtering-based algorithms, the best performance for male users is achieved when the recommender captures the attractiveness of recommended users to the service user and interest from the service user in recommended users. On the contrary, the best performance for females is achieved when recommended users are interested in the service user and the service user is attractive to recommended users.  The results show that when looking for potential dates, males tend to be focused on their own interest and oblivious towards their attractiveness to potential dates, while females are more conscientious to their own attractiveness to and interest from the other side on the line.

\end{itemize}

The rest of the paper is organized as follows. Section \ref{sec:related_work} describes the related work on reciprocal recommendation as well as reciprocal relation prediction in online social networks. Section \ref{sec:algorithm} presents the reciprocal recommendation algorithms we proposed. Description and characteristics of our dataset are provided in Section \ref{sec:dataset}. Section \ref{sec:evaluation} presents the performance of our proposed algorithms. Finally, we conclude our paper in Section \ref{sec:conclusion}.
\section{Related Work} \label{sec:related_work}
A few studies on the analysis of user behavior of online dating sites have provided valuable guidelines to design recommendation system 
for online dating. In \cite{Peng2014}, the authors analyze how user's sending and replying behavior correlate with several important user 
attributes, such as age, income, education level, and number of photos, etc., and how much a user's actual preference deviates 
from his/her stated preference. The findings also correspond to the result of \cite{Luiz2010Euro} that the recommendation system built 
on user's implicit preference outperforms that built on user's explicit preference.

There exists research that aims to predict user reciprocity in various online social networks. In \cite{Junichiro2012}, a machine learning based approach is proposed to find plausible candidates of business partners using firm profiles and transactional relations between them. The authors of \cite{John2011} propose a Triad Factor Graph model to deal with two-way relation prediction in online social networks. In \cite{PengXia2014}, both user-based and graph-based features are applied in a machine learning framework to predict user replying behavior in online dating networks. A new 
collaborative filtering approach called \textit{CLiMF} \cite{Yue2012} is proposed to learn latent factors of users and items and improves the performance of top-k recommendations for recommending friends or trustees on Epinions and Tuenti. Further, they improve the algorithm by optimizing Expected Reciprocal Rank, an evaluation metric that generalizes reciprocal rank in order to incorporate user feedback with multiple levels of relevance in \cite{Yue2013}. 

There have been recently several studies on the people to people recommendation for online social networks. Both content based algorithm and collaborative filtering method are applied to recommend users to follow in Twitter \cite{John2010}. A LDA-based method is employed in \cite{Gang2013} to discover communities in Twitter-style online social networks, and matrix factorization are applied on each community to provide recommendations.

For online dating recommendations, the authors in  \cite{Lei2012} and \cite{Luiz2010} analyzed the characteristics of reciprocal recommendations in detail. In particular, \cite{Lei2012} considers both local utility (users' mutual preference) and global utility (overall bipartite network), and proposes a generalized reciprocal recommendation framework for both online dating sites and online recruiting sites.  A content based reciprocal  algorithm (RECON) proposed in \cite{Luiz2010} learns the preference of both sides of users and defines a new evaluation metric (success rate) to evaluate the performance of their algorithm. In their following work \cite{Luiz2011,LuizA2011}, RECON is extended to consider both positive and negative preference, and collaborative filtering is applied with stochastic matching algorithm. In \cite{Kang2013}, a hybrid collaborative filtering based algorithm that takes reciprocal links into account is proposed and shown to have good performance in recommending both initial and reciprocal contacts. A recent study of  \cite{Joshua011} finds that users with similar profiles like and dislike similar people, and are liked and disliked by similar people. This hypothesis is used to build a content-collaborative reciprocal recommender, which is evaluated on a popular online dating dataset. In \cite{Xiongcai2010}, collaborative filtering algorithms are used to capture the reciprocity in people to people recommendation. The authors in \cite{Kun2014} propose a two-side matching framework for online dating recommendations and design a Latent Dirichlet Allocation (LDA) model to learn the user preferences from the observed user messaging behavior and user profile features. 

The studies most relevant to our online dating recommendation problem are \cite{Luiz2010} and \cite{Kang2013}, in which a content-based algorithm (RECON) and a collaborative filtering-based algorithm (HCF) was presented and shown to outperform other approaches. In this paper we will compare the performance of our proposed algorithms with these two algorithms.

\section{Recommendation System} \label{sec:algorithm}
In this section, we propose a generalized reciprocal recommendation system that aims to match people with mutual interest in each other. We also introduce two previously proposed approaches, namely, a content-based algorithm RECON \cite{Luiz2010} and a hybrid collaborative filtering algorithm (HCF) \cite{Kang2013} for comparison.

\subsection{System Design}
The success of a reciprocal recommendation system lies in its ability to recommend users with whom the service user has mutual interest and thus they are likely to communicate with each other. The interaction records between a pair of users are a good indicator of actual interest and attractiveness between the sender and receiver. If a recommended user matches the service user's interest, the service user will be more likely to approach the recommended user. Also, if the service user's attractiveness is compatible with the recommended user's interest, the recommended user will be more likely to reply to the service user when contacted. 


\begin{figure}[htb]
\centering
\includegraphics[width=0.8\textwidth]{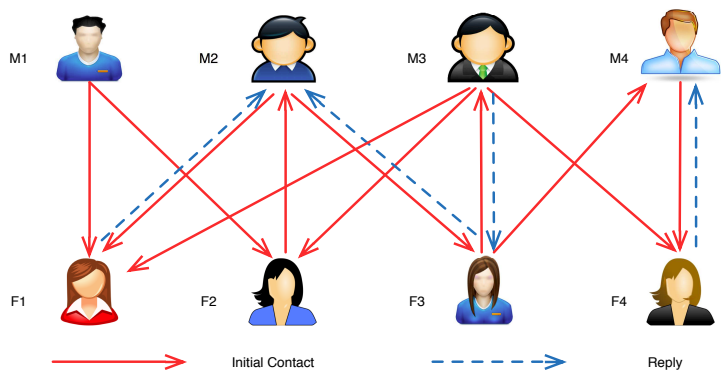}
\caption{Example of an online dating recommendation problem}
\label{fig:example}
\end{figure}

Figure \ref{fig:example} depicts an example of an online dating network. Based on user attributes and their communication traces, our goal is to match users with mutual interest in each other, for example,  M3 and F3, M4 and F4. In the following we will describe how to measure the similarity between a pair of users in terms of their dating interest and attractiveness, and how to construct various recommendation algorithms based on these similarity measures.

Our reciprocal recommendation system is comprised of the following four major components.

\vspace{10pt}
\noindent \textbf{Extracting User Based Features}: When a user registers with the online dating site, he/she needs to provide a variety of profile information including the user's gender, age, current location (city level), home town location, height, weight, body type, blood type, occupation, income range, education level, religion, astrological sign, marriage and children status, photos, home ownership, car ownership, interests, smoking and drinking behavior. Most of these attributes are categorical features, except age, height, weight and number of photos. 

\begin{algorithm}[H]
	\caption{Reciprocal Score($x$, $y$)}
	\label{alg:reciproal-algorithm}
	\begin{algorithmic}
		
		\State \textbf{Input}: service user $x$, and recommended user $y$
		\State \textbf{Output}: reciprocal score 
		\State \textbf{begin}
		
		\State \hspace{\algorithmicindent} /* initialize compatible scores s*/
		\State \hspace{\algorithmicindent} $s(x, y) = 0.0$
		\State \hspace{\algorithmicindent} $s(y, x) = 0.0$ 
		\State \hspace{\algorithmicindent} /* compute compatible scores for both $x$ and $y$ */
		\State \hspace{\algorithmicindent} \textbf{foreach} $u$ in $Neighbor_{1}(y)$:
		\State \hspace{\algorithmicindent} \hspace{\algorithmicindent} $s(x, y) = s(x, y) + Similarity_{1}(x, u)$
		\State \hspace{\algorithmicindent} \textbf{end for}
		\State \hspace{\algorithmicindent} \textbf{foreach} $v$ in $Neighbor_{2}(x)$:
		\State \hspace{\algorithmicindent}  \hspace{\algorithmicindent} $s(y, x) = s(y, x) + Similarity_{2}(y, v)$
		\State \hspace{\algorithmicindent} \textbf{end for}
		\State \hspace{\algorithmicindent} /* normalize compatible scores for both $x$ and $y$ */
		\State \hspace{\algorithmicindent} if $ |Neighbor_{1}(y)| > 0 $
		\State \hspace{\algorithmicindent} \hspace{\algorithmicindent} $s(x, y) = \frac{s(x, y)}{|Neighbor_{1}(y)|}$ 
		\State \hspace{\algorithmicindent} if $ |Neighbor_{2}(x)|  > 0 $
		\State \hspace{\algorithmicindent} \hspace{\algorithmicindent} $s(y, x) = \frac{s(y, x)}{|Neighbor_{2}(x)|}$ 
		\State \hspace{\algorithmicindent} /* compute reciprocal score */
		\State \hspace{\algorithmicindent} if  $s(x, y) > 0$ and $s(y, x) > 0$
		\State \hspace{\algorithmicindent} \hspace{\algorithmicindent} return $\frac{2}{(s(x, y))^{-1}+(s(y, x))^{-1}}$
		\State \hspace{\algorithmicindent} else
		\State \hspace{\algorithmicindent} \hspace{\algorithmicindent} return $0.0$
		
		\State \textbf{end}
	\end{algorithmic}
\end{algorithm}

\vspace{10pt}
\noindent \textbf{Extracting Graph Based Features}:
The online dating site we study is for heterosexual dating and only allows communications between users of opposite sex. The communication traces between users can be modeled as a bipartite network between males and females. A collection of graph-based similarity features are extracted from 
the bipartite graph that represent a user's active level in dating and attractiveness. The detailed definitions of these graph based features are provided in Section \ref{sec:similarity_function}.
   
\vspace{10pt}
\noindent \textbf{Compute Reciprocal Scores}:
Based on both user-based and graph-based features, we use the reciprocal score to measure the mutual
interest and attractiveness between a pair of potential dates as described in Algorithm \ref{alg:reciproal-algorithm}.

In Algorithm \ref{alg:reciproal-algorithm}, $Neighbor_{1}$() and $Neighbor_{2}$() represent the
neighbor set of a user with different directions in the bipartite network, and their formal definitions are
given in equations (\ref{eq:g1}) and (\ref{formula:graph}). $Similarity_{1}(,)$ and $Similarity_{2}$(,) are customizable functions 
measuring the similarity between two users. We will discuss these functions in the following subsection. 
Given $Neighbor_{1}$, $Neighbor_{2}$, $Similarity_{1}$ and $Similarity_{2}$, $s(x, y)$  
measures the similarity between user $x$ and user $y$'s neighbors, while $s(y,x)$ measures the similarity between
user $y$ and user $x$'s neighbors.  After computing $s(x, y)$ and $s(y, x)$, the reciprocal score is computed as 
the harmonic mean of these two similarity scores. 

\vspace{10pt}
\noindent \textbf{Generate Recommended User List}:
For a given service user, a recommendation list will be generated by ranking these reciprocal scores. 
We will present the top-K users in the list to the service user. Note that the reciprocal 
score may not be symmetric if $Neighbor_{1}$ and $Neighbor_{2}$ are set as different functions or 
$Similarity_{1}$ and $Similarity_{2}$ are set as different functions. This is different from the case 
in RECON where there is a unique reciprocal score for any pair of users.

\subsection{Similarity Functions} \label{sec:similarity_function}
\subsubsection{Content-based Similarity Functions}
In content-based recommendation system, every recommended user can be represented by a feature vector or an 
attribute profile. These features can be numeric or nominal values representing some aspect of the user, such 
as age, height, income, and etc. 
Let $A_{x}$ denote the set of known attributes (age, height, income, education level, etc) of 
user $x$, i.e.,
\begin{equation*}
  A_{x} = \{a: \mbox{$a$ is a known attribute of user $x$}\}
\end{equation*}

We denote the set of user-based attributes of user $x$ as 
\begin{equation}
	U_{x} = \{v^{x}_{a}: \mbox{ for $a$ $\in$ } A_{x}\},
\end{equation}
where $v^{x}_{a}$ is the value of attribute $a$ of user $x$.

The first similarity measure based on user attributes follows the work of RECON \cite{Luiz2010}, where the values of the numeric attributes (e.g., age, height, and etc) are grouped into ranged nominal values. For each service user, RECON builds his/her preference model by constructing the distribution of the receivers' user attributes. How much a service user is interested in a recommended user is then measured by comparing the attributes of the recommended user with the preference model of the service user. This is equivalent to computing the similarity between two users as follows:
\begin{equation}
	\begin{aligned}
	\mbox{content-similarity}^{a}(x, y) = \frac{\sum_{a \in A_{x} \cap A_{y} }{P_{a}(x, y)}}{|A_{x} \cap A_{y}|}
	\end{aligned}
	\label{formula:recon}
\end{equation}
where 
\begin{equation}
P_{a}(x, y) = \left\{
             \begin{array}{lcl}
             {1,} &\text{if $v^{x}_{a} = v^{y}_{a}$}  \\
             {0,} &\text{otherwise.} 
             \end{array}  
        \right.
\end{equation}

In RECON, numerical values are converted into categorical values, for example, users are divided into age groups 20-24, 25-29, 30-34, etc. Note that this method does not capture the numerical attribute information at the boundaries of continuous ranges. For example,  a user of age 25 will not be considered to be similar to a user of age 24 as they fall into different age ranges. To avoid the loss of information, we modify the similarity measure defined in equation (\ref{formula:recon}) as follows and use it as our second similarity measure, i.e.,
\begin{equation}
	\begin{aligned}
	\mbox{content-similarity}^{b}(x, y) = \frac{\sum_{a \in A_{x} \cap A_{y} }{Q_{a}(x, y)}}{|A_{x} \cap A_{y}|},
	\end{aligned}
	\label{formula:content}
\end{equation}
where $Q_{a}(x, y) = P_{a}(x, y)$ for the nominal attributes as before; for numeric attributes, we define 
\begin{equation}
	\begin{aligned}
	Q_{a}(x, y) = \frac{v^{*}_{a}-|v^{x}_{a} - v^{y}_{a}|}{v^{*}_{a}},
	\end{aligned}
\end{equation} 
where $v^{*}_{a} = \max_{i\neq j} |v_{a}^{i} - v_{a}^{j}|$ represents the maximum absolute difference for attribute $a$ among all users. This new similarity measure results in a value between 0 when the attributes of the two users have the maximum difference (i.e., $|v_{a}^{i} - v_{a}^{j}| = v^{*}_{a}$) and 1 when the attributes of the two users are the same (i.e., $v_{a}^{i} = v_a^{j}$). For the above example, the age similarity between a 24-year-old user and a 25-year-old user will be 48/49 (very close to 1), where 49 is the max difference in user ages. It is clear that this new measure can better capture the similarity of numerical attributes between two users than the measure defined in equation (\ref{formula:recon}), and as will be shown in Section \ref{sec:evaluation}, results in better performance in generating the recommendation list.

The above two similarity measures are based on user attributes, and can be used to construct variations of content-based recommendation algorithms. 

\subsubsection{Graph-based Similarity Functions}
Based on the message traces between users, we define the following two graph-based measures to capture the user's active level and attractiveness:
\begin{eqnarray}
	Se(x) & = & \{y: \mbox{$x$ has sent a messages to $y$} \} \label{eq:g1}\\	
	Re(x) & = & \{y: \mbox{$x$ has received a message from $y$} \} 
	\label{formula:graph}
\end{eqnarray}
where $Se(x)$ is defined as the set of out-neighbors of $x$ and its cardinality reflects the activeness of 
user $x$; $Re(x)$ is defined as the set of in-neighbors of $x$ and its cardinality reflects the 
attractiveness of user $x$. 

Based on the graph-based measures defined in equations (\ref{eq:g1}) and (\ref{formula:graph}), we introduce the following two similarity functions:
\begin{itemize}
\item  {\it Interest similarity}: Consider two users of the same gender, $x$ and $y$. If they both contact a same user, it shows that they share common interest in looking for potential dates. The fraction of users who received messages from both $x$ and $y$ among all users who received messages from either $x$ or $y$ serves as a measure of the similarity between the dating interest of the two users, i.e.,
\begin{equation}
	 \mbox{interest-similarity}(x, y) = \frac{|Se(x) \cap Se(y)|}{|Se(x) \cup Se(y)|},
	\label{formula:send-collab}
\end{equation}

Note that we adopt the Jaccard Coefficient in the interest similarity measure as the number of shared receivers from a pair of users is typically far outnumbered by the total number of receivers from the two users. This will become clear based on the degree distribution of users shown in Figure \ref{fig:messageCCDF} and weight distribution of the projection network shown in Figure \ref{fig:sendProjection} from the dataset description in Section \ref{sec:dataset}. For the example shown in Figure \ref{fig:example}, together M1 and M2 contacted three different females among which F2 received messages from both of them. The interest-similarity between M1 and M2 is thus 1/3.

\item {\it Attractiveness similarity}: Consider two users of the same gender, $x$ and $y$. If they both receive messages from a same user, it shows that they share common attractiveness to potential dates. The fraction of users who sent messages to both $x$ and $y$ among all users who sent messages to either $x$ or $y$ serves as a measure of the similarity between the attractiveness of the two users, i.e.,
\begin{equation}
	\mbox{attractiveness-similarity}(x, y) = \frac{|Re(x) \cap Re(y)|}{|Re(x) \cup Re(y)|}.
	\label{formula:receive-collab}
\end{equation}

By the same token we adopt the Jaccard Coefficient in the attractiveness similarity measure. For the example shown in Figure \ref{fig:example}, both F1 and F2 received messages from M1 and M3 while F1 also received messages from M2. The attractiveness similarity between F1 and F2 is thus 2/3.

\end{itemize}

\subsection{Recommendation Algorithms}

Based on these four similarity functions, we construct the following two  content based algorithms and four collaborative filtering algorithms.

The content based algorithms are constructed based on the  similarity of user attributes, including:
\begin{itemize}
  \item \textbf{CB1 (RECON)}: Both  $Neighbor_{1}$ and $Neighbor_{2}$ in Algorithm \ref{alg:reciproal-algorithm} are set as out-neighbors $Se()$,  and both $Similarity_{1}$ and $Similarity_{2}$ are computed using content similarity defined in equation (\ref{formula:recon}). This algorithm is the same as RECON \cite{Luiz2010}.

  \item \textbf{CB2}: Both  $Neighbor_{1}$ and $Neighbor_{2}$ are set as out-neighbors $Se()$, and both $Similarity_{1}$ and $Similarity_{2}$ are computed using content similarity function  defined in equation (\ref{formula:content}), where we do not convert numeric attributes into nominal attributes.
\end{itemize}

\begin{figure}
\subfloat[\label{fig:AS}]
  {\includegraphics[width=.5\linewidth]{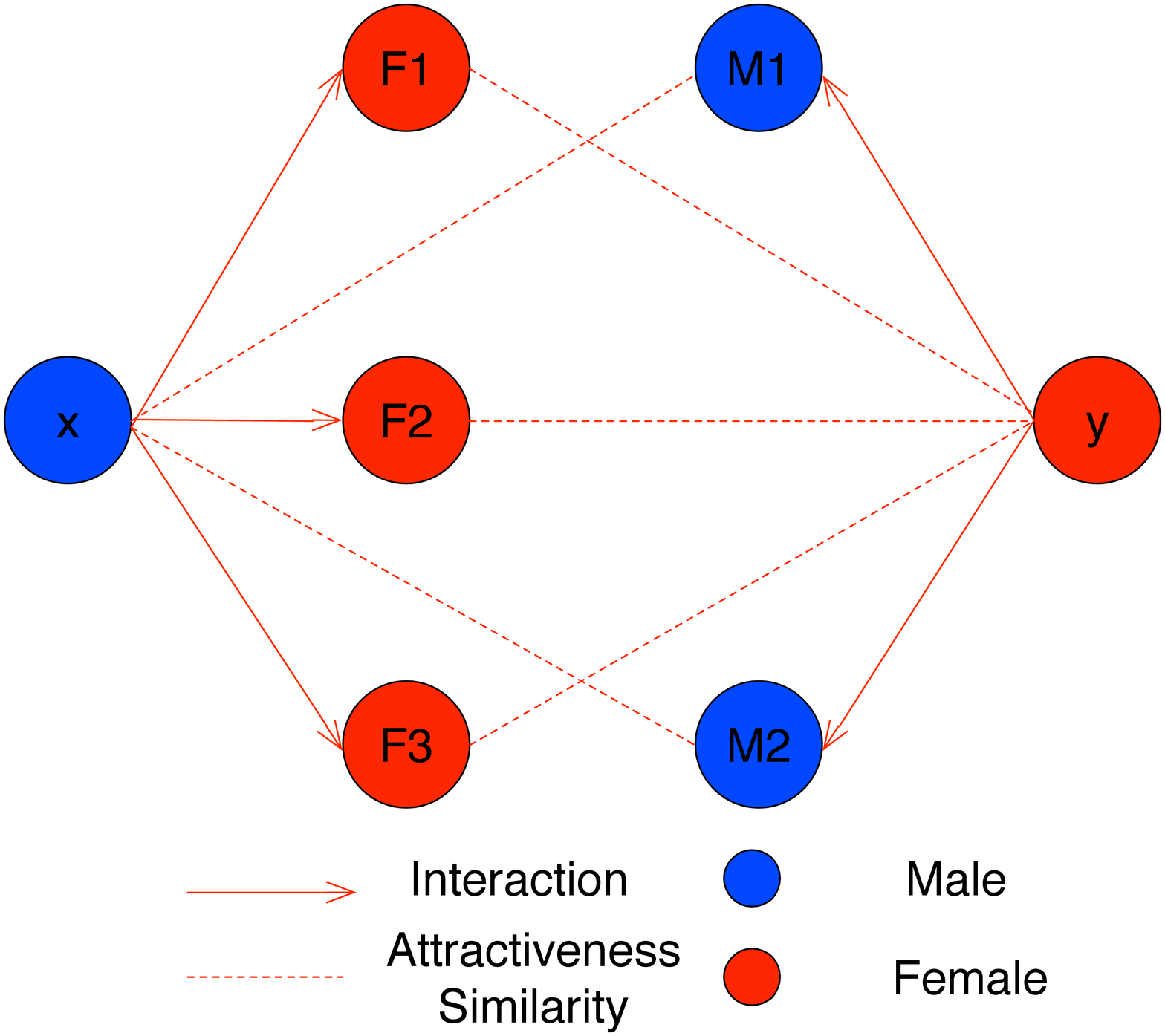}}\hfill
\subfloat[\label{fig:IS}]
  {\includegraphics[width=.5\linewidth]{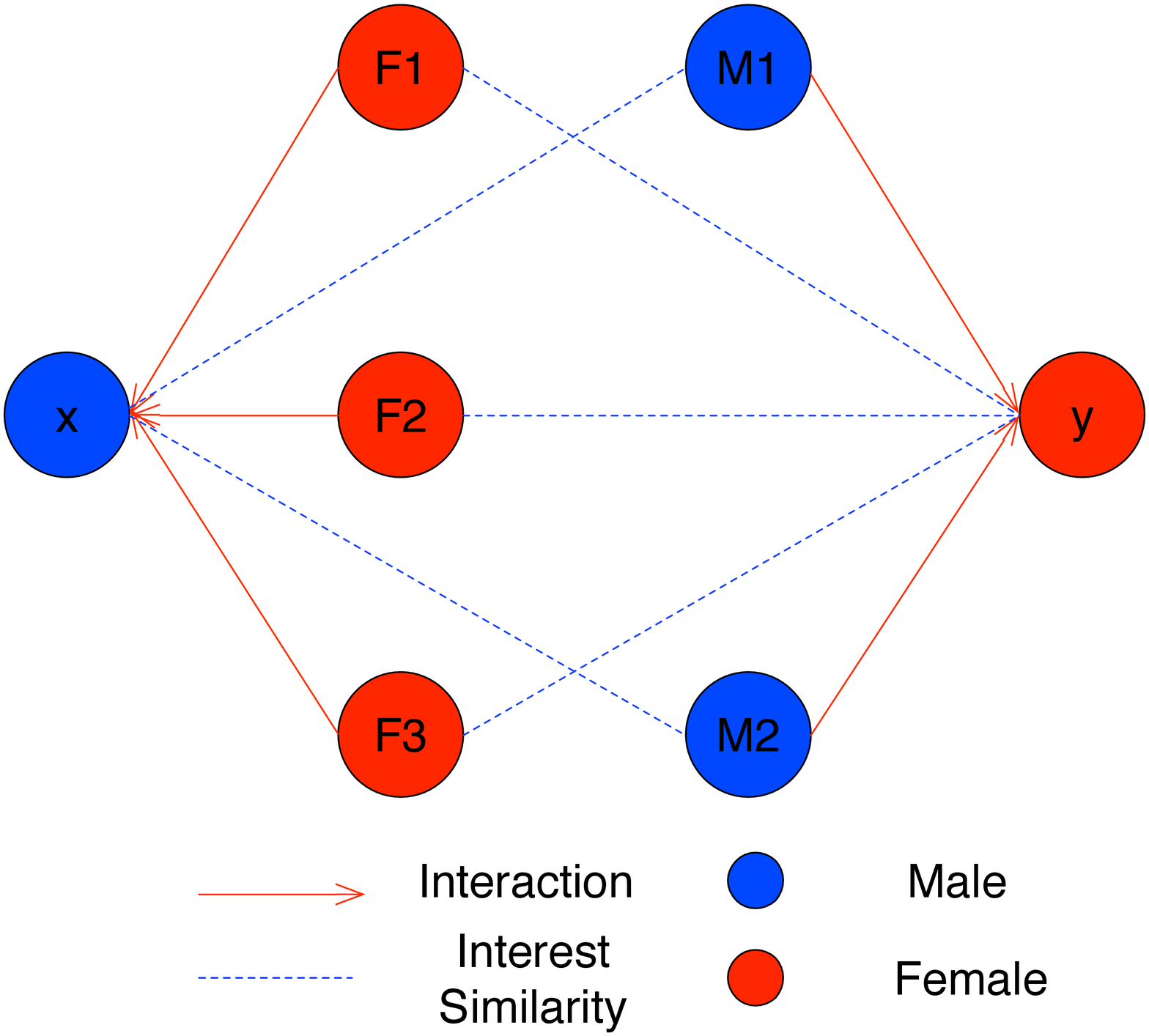}}\hfill
\caption{Example of (a)CF1 and (b)CF2 algorithms.}
\label{fig:algorithm_example}
\end{figure}

\begin{figure}
\subfloat[\label{fig:CF3}]
  {\includegraphics[width=.5\linewidth]{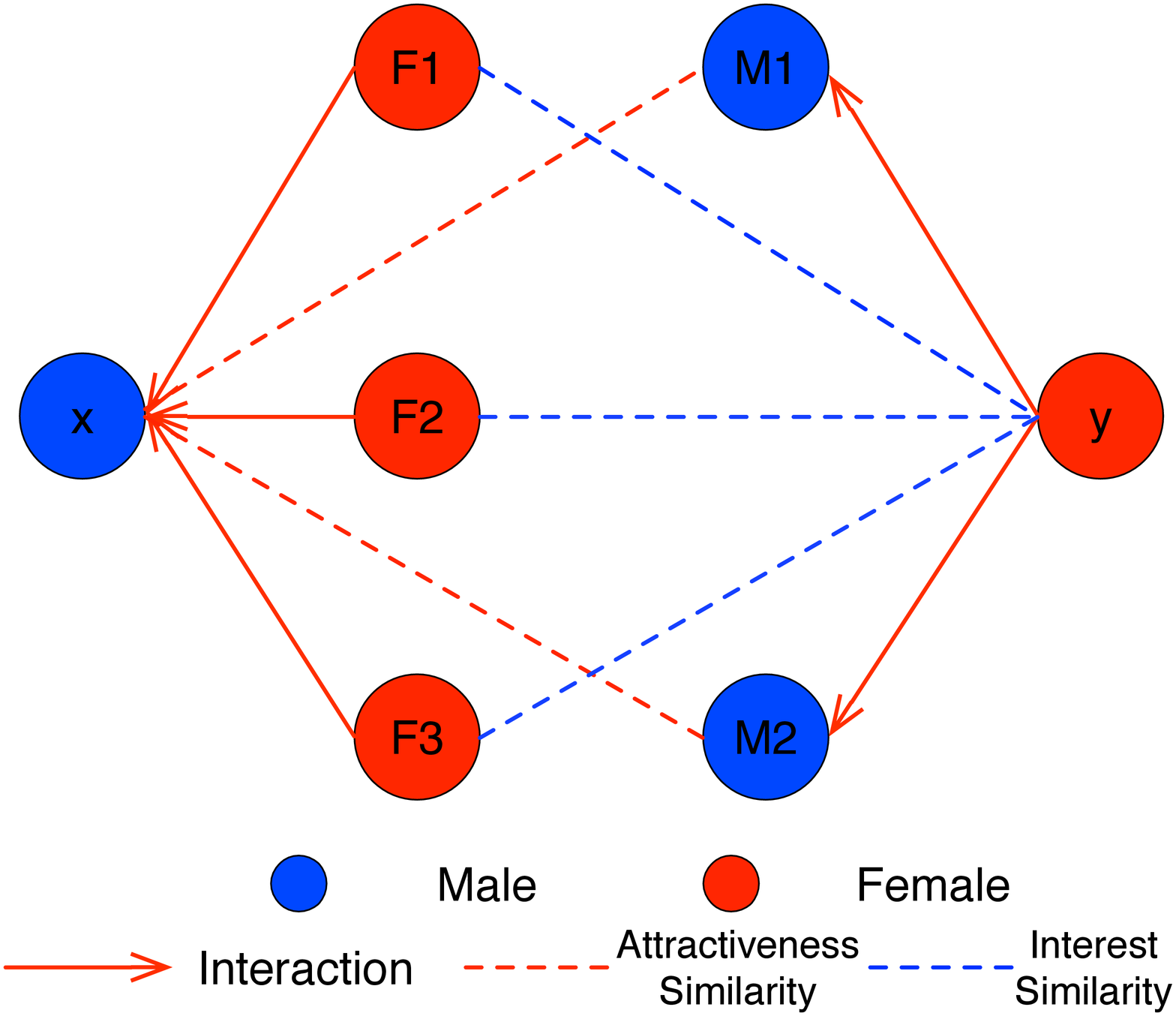}}\hfill
\subfloat[\label{fig:CF4}]
  {\includegraphics[width=.5\linewidth]{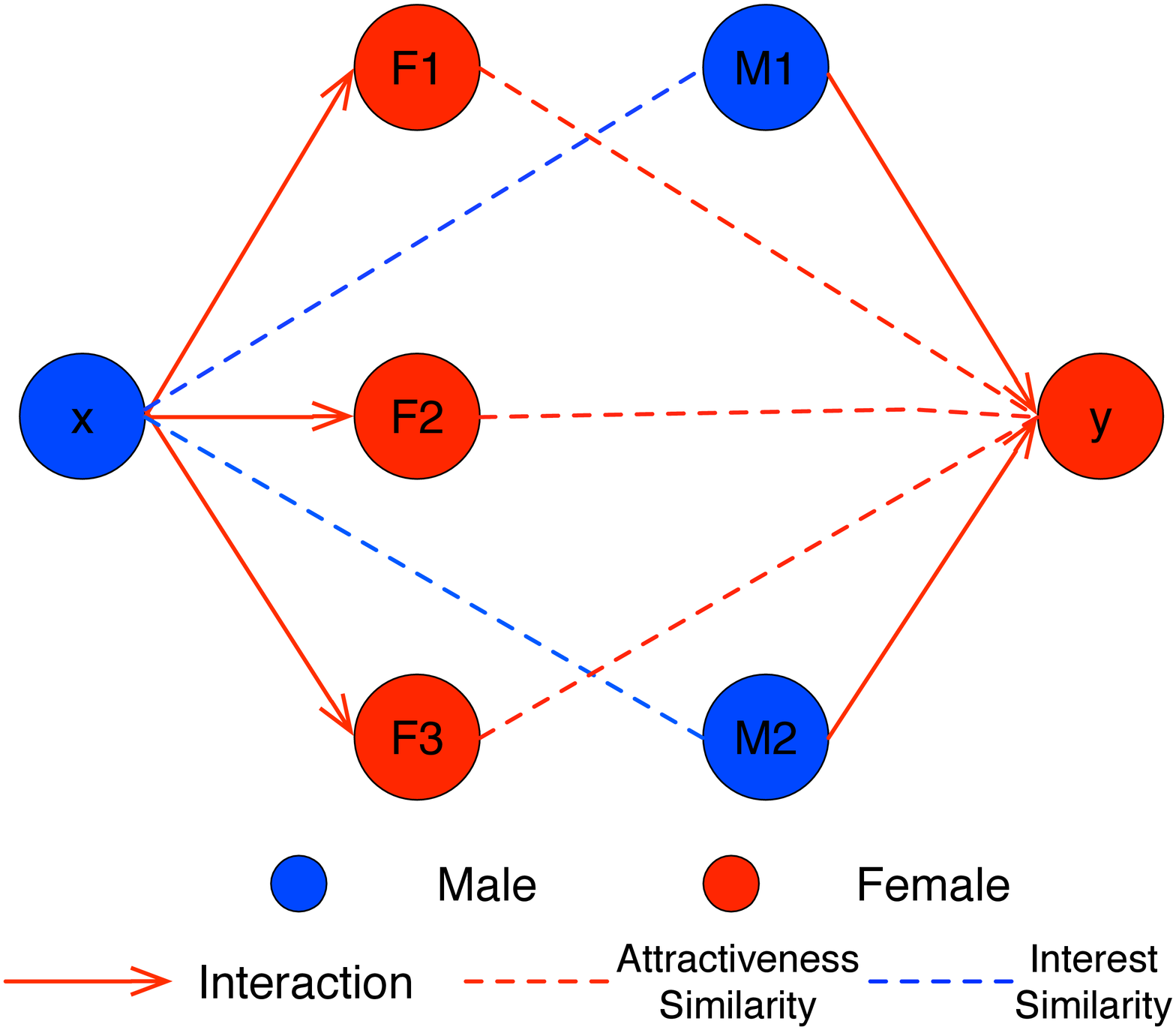}}\hfill
\caption{Example of (a)CF3 and (b)CF4 algorithms.}
\label{fig:algorithm_example_2}
\end{figure}

Collaborative filtering-based algorithms make use of the communication history of the service user as well as the decisions made by users with similar interest or attractiveness to help make recommendations. Based on different combinations of users' dating interest and attractiveness, we construct the following four collaborative filtering-based recommendation algorithms:

\begin{itemize}
  \item \textbf{CF1}: Both  $Neighbor_{1}$ and $Neighbor_{2}$ are set as out-neighbors $Se()$, and both $Similarity_{1}$ and $Similarity_{2}$ are computed using attractiveness similarity defined in equation (\ref{formula:receive-collab}). Therefore, for Algorithm \ref{alg:reciproal-algorithm} we have 
  \begin{eqnarray*}
  s(x,y)= \frac{1}{|Se(y)|}\sum_{k\in Se(y)}\mbox{attractiveness\_similarity}(x,k) \\
  s(y,x)=\frac{1}{|Se(x)|}\sum_{k\in Se(x)}\mbox{attractiveness\_similarity}(k,y)
  \end{eqnarray*}
  
  In this case, $s(x,y)$ sums up the attractiveness similarity (i.e., contacted by same users) between service user $x$ and users who have been contacted by user $y$, capturing the attractiveness of the service user $x$ to user $y$. Similarly, $s(y,x)$ captures the attractiveness of user $y$ to service user $x$. Putting these two factors together, this algorithm captures the mutual attractiveness between the service user and recommended users. An example of CF1 algorithm is shown in Figure \ref{fig:AS}.

  \item \textbf{CF2}: Both  $Neighbor_{1}$ and $Neighbor_{2}$ are set as in-neighbors $Re()$. Both $Similarity_{1}$ and $Similarity_{2}$ are computed using interest similarity defined in equation (\ref{formula:send-collab}). In this case, $s(x,y)$ sums up the interest similarity between service user $x$ and users who have contacted user $y$, capturing the interest from user $x$ to $y$. Similarly, $s(y,x)$ captures the interest from user $y$ to $x$.
  Together, this algorithm captures the mutual interest between the service user and recommended users. An example of CF2 algorithm is shown in Figure \ref{fig:IS}.

  \item \textbf{CF3}: $Neighbor_{1}$ is set as out-neighbors $Se()$, while $Neighbor_{2}$ is set as out-neighbors $Re()$. $Similarity_{1}$ is computed using attractiveness similarity defined in equation (\ref{formula:receive-collab}), and $Similarity_{2}$ is computed using interest similarity defined in equation (\ref{formula:send-collab}). This algorithm captures the interest from recommended users in the service user and the attractiveness of the service user to recommended users. An example of CF3 algorithm is shown in Figure \ref{fig:CF3}. 

  \item \textbf{CF4}: $Neighbor_{1}$ is set as in-neighbors  $Re()$ , while $Neighbor_{2}$ is set as out-neighbors $Se()$.  $Similarity_{1}$ is computed using interest similarity defined in equation (\ref{formula:send-collab}), and $Similarity_{2}$ is computed using attractiveness similarity defined in equation (\ref{formula:receive-collab}). This algorithm captures the attractiveness of recommended users to the service user and the interest of the service user in recommended users. An example of CF4 algorithm is shown in Figure \ref{fig:CF4}.
  


\end{itemize}

Figure \ref{fig:algorithm_example} illustrates the mechanism of CF1 and CF2 algorithms. In the example of CF1 shown in Figure \ref{fig:AS}, service user $x$ sent messages to users $F1$, $F2$ and $F3$. These out-neighbors of service user $x$ share similar attractiveness with user $y$, i.e., user $y$ and $F1$, $F2$, $F3$ received messages from at least one common user. Also, user $y$ sent messages to users $M1$ and $M2$ who share similar attractiveness of user $x$. Therefore, CF1 captures the mutual attractiveness between the service user and recommended users. If the reciprocate score between $x$ and $y$ ranks in the top-K position, user $y$ will be included in the recommendation list for service user $x$. Figure \ref{fig:IS} shows an example of CF2, which captures the mutual interest between the service user $x$ and user $y$, i.e., user $x$'s interest in user $y$ and user $y$'s interest in service user $x$. Examples of CF3 and CF4 illustrated in Figure \ref{fig:algorithm_example_2} can be interpreted in a similar way. 

In addition to the above algorithms, we also implement the content-based algorithm (RECON) proposed in \cite{Luiz2010} and the hybrid collaborative filtering algorithm (HCF) proposed in \cite{Kang2013}. In particular, RECON corresponds to our CB1 algorithm, where the $Neighbor_{1}$ and $Neighbor_{2}$ are set as $Se()$ function, and $Similarity_{1}$ and $Similarity_{2}$ are computed based on equation (\ref{formula:recon}). HCF extends the baseline collaborative filtering approach by considering both initial and reciprocal contacts to compute the similarity between two users, where reciprocal links are given higher weight than single direction contacts. These two algorithms are most related to our study and have been shown to outperform many other approaches. We will compare the performance of our proposed algorithms with these two algorithms in Section \ref{sec:evaluation}.
 
\section{Dataset Description} \label{sec:dataset}

\subsection{Dataset Overview}

The dataset used in our study is obtained through a collaboration with baihe.com, one of the major online 
dating sites in China. Our dataset includes the profile information of 200,000 users uniformly sampled from users 
registered in November of 2011. Of the 200,000 sampled users, 139,482 are males and 60,518 are females, 
constituting 69.7\% and 30.3\% of the total number of sampled users respectively.
 For each user, we have his/her message sending and receiving traces (who contacted whom at what 
time) in the online dating site and the profile information of the users that he or she has communicated 
with from the date that the account was created until the end of January 2012. Note that the site is 
for heterosexual dating and only allows communications between users of opposite sex.

After a user creates an account on the online dating site, he/she can search for potential dates based 
on information within the profiles provided by the other users including user location, age, etc. Once 
a potential date has been discovered, the user then sends a message to him/her, which may or may not be 
replied by the recipient. In this paper we focus on the prediction of whether a user will reply to 
initial messages sent by other users. Subsequent interactions between the same pair of users do not 
represent a new sender-receiver pair and can not be used as the only indicator for continuing 
relationship as users may choose to go off-line from the site and communicate via other channels 
(e.g., email, phone or meet in person).

\begin{table}[htb]
\caption{ Dataset Description}
\centering
\scalebox{1}{
\begin{tabular}{p{3.3cm} | p{2cm} | p{2.3cm} }
\hline\hline
Type & Initial contact links & Reciprocal links (Reply rate) \\
\hline
Male to Sample Female & 1,586,059 & 150,917 (9.5\%) \\
Female to Sample Male & 328,645 & 58,946 (17.9\%)\\
\hline
\end{tabular}
}
\label{table:dataset}
\end{table}

Since we only have eight full weeks' worth of online dating interaction records for our sample users, 
we will consider the activities of each user during the first eight weeks of his/her membership. 
Table \ref{table:dataset} describes the characteristics of the dataset. More detailed description and 
analysis of the dataset can be found in our recent work \cite{Peng2013,Peng2014}.

\begin{figure}
\subfloat[\label{fig:MessageSentCCDF}]
  {\includegraphics[width=.5\linewidth]{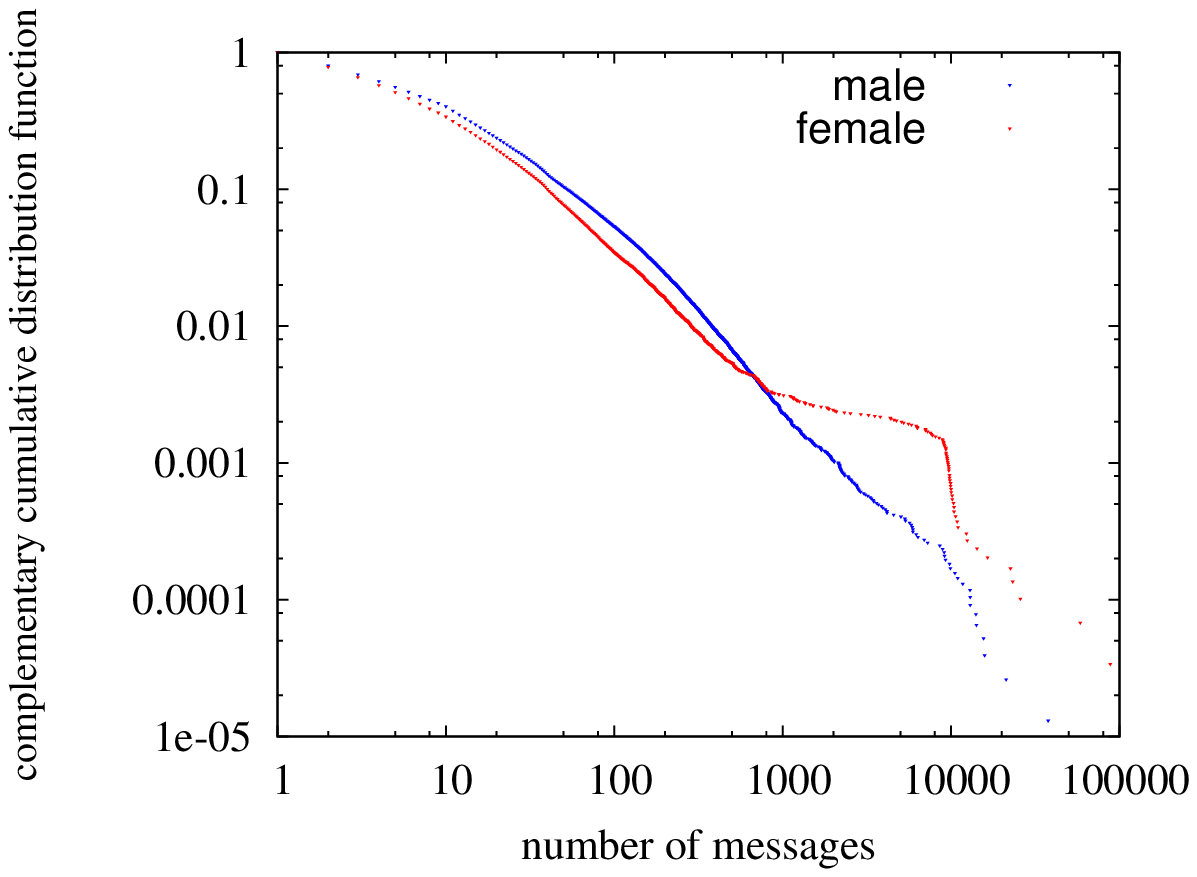}}\hfill
\subfloat[\label{fig:MessageReceivedCCDF}]
  {\includegraphics[width=.5\linewidth]{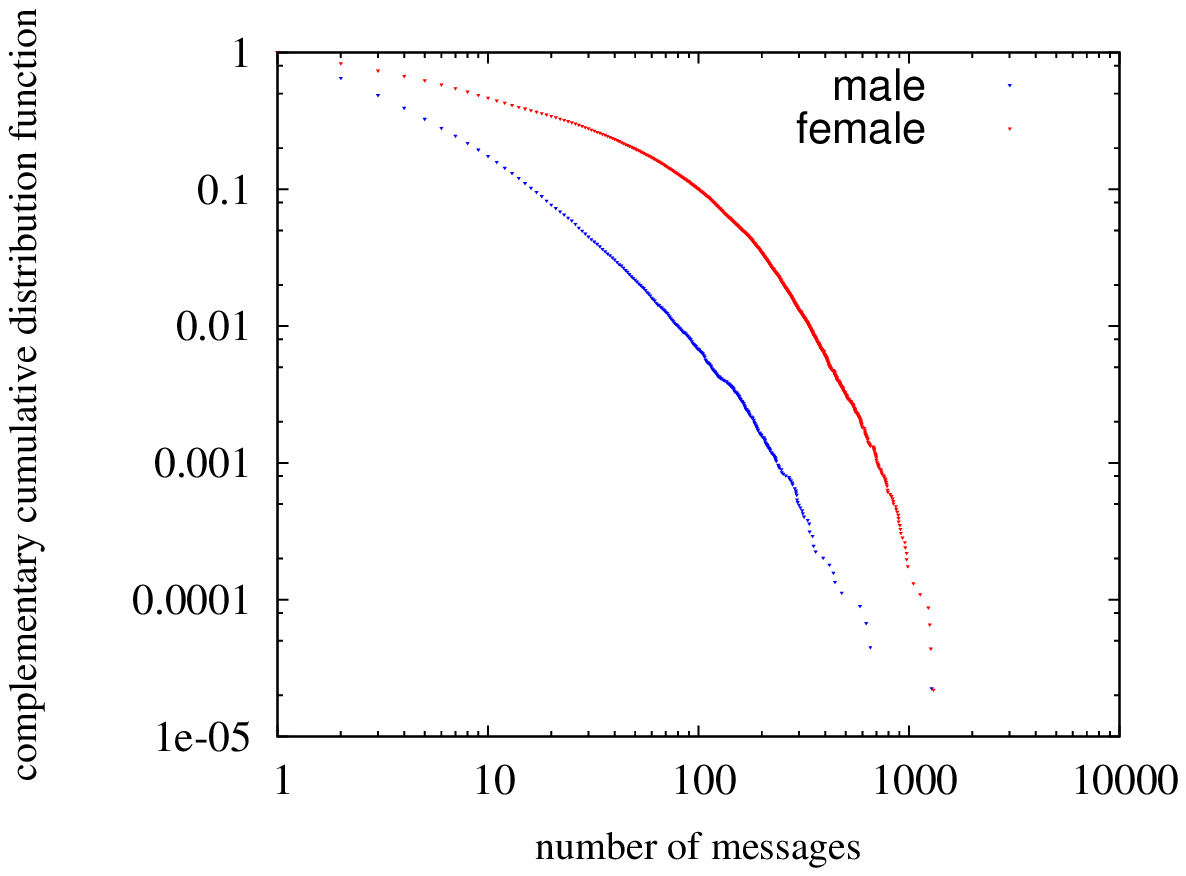}}\hfill
\caption{(a) CCDF of the number of messages a user sent out during the first eight weeks of his/her membership. (b) CCDF of the number of messages a user received during the first eight weeks of his/her membership.}
\label{fig:messageCCDF}
\end{figure}

For both males and females, we obtain the distribution of the number of messages sent by each user per week given that a user sends at least one message during the week, and plot its complementary cumulative density function (CCDF) in Figure \ref{fig:messageCCDF}\subref{fig:MessageSentCCDF}. We observe that the distributions exhibit  heavy tails. Most users only sent out a small number of messages: 94.6\% of males and 96.5\% of females  sent out less than 100 messages during the first eight weeks of their membership. On the other hand, there are small fractions of users that sent out a large number of messages. According to the online dating site, most of these highly active users are likely to be fake identities created by spammers and their accounts have been quickly removed from the site. 

The distribution of number of messages received by a user is plotted in Figure \ref{fig:messageCCDF}\subref{fig:MessageReceivedCCDF}. A female is likely to receive more messages than a male. Most users only received a small number of messages: 99.3\% of males and 90.1\% of females received less than 100 messages during the first eight weeks of their membership. On average, a male received 7 messages while a female received 35 messages during the first eight weeks.

\begin{figure}
\subfloat[\label{fig:maleSend}]
  {\includegraphics[width=.5\linewidth]{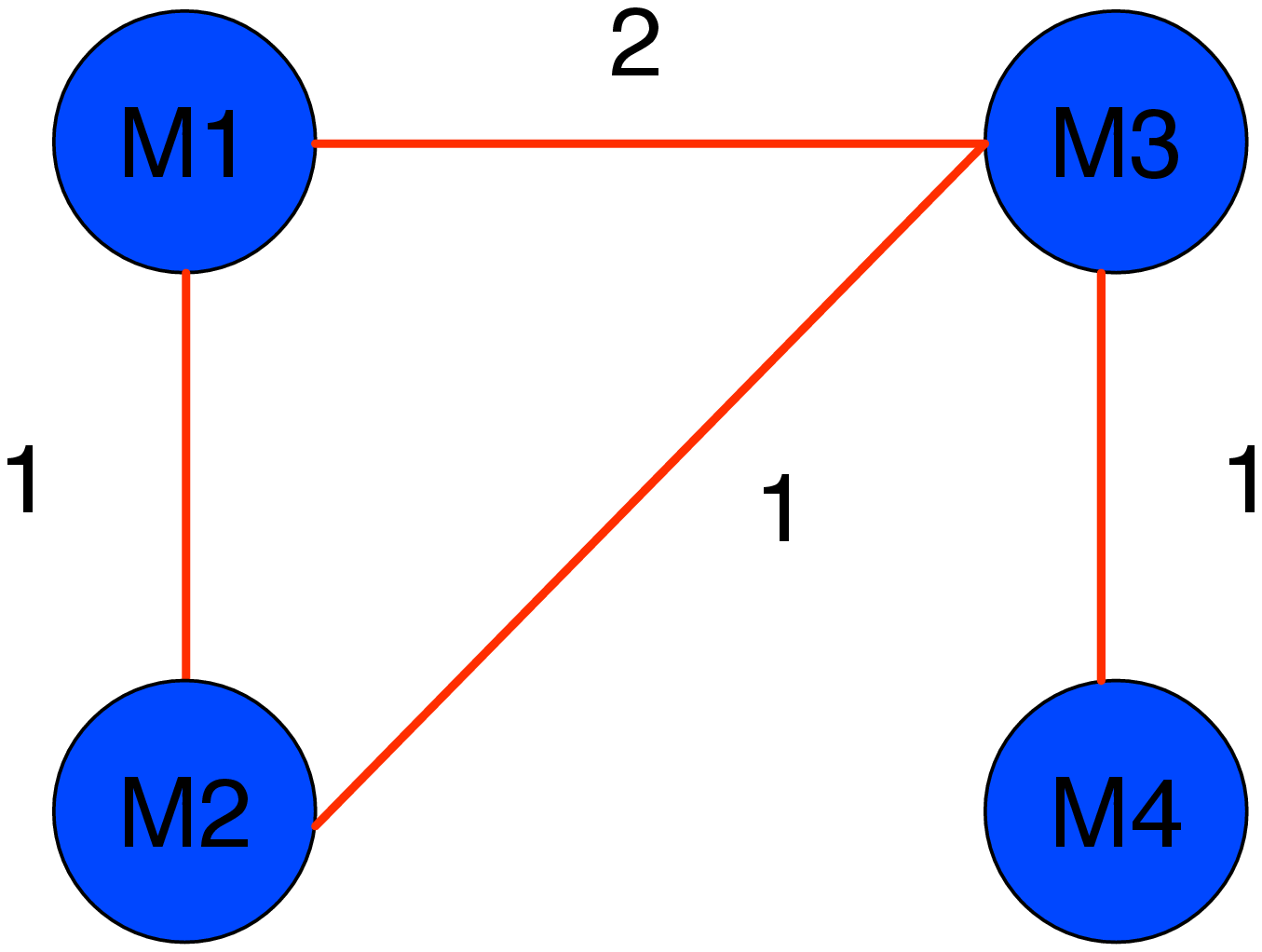}}\hfill
\subfloat[\label{fig:femaleReceive}]
  {\includegraphics[width=.5\linewidth]{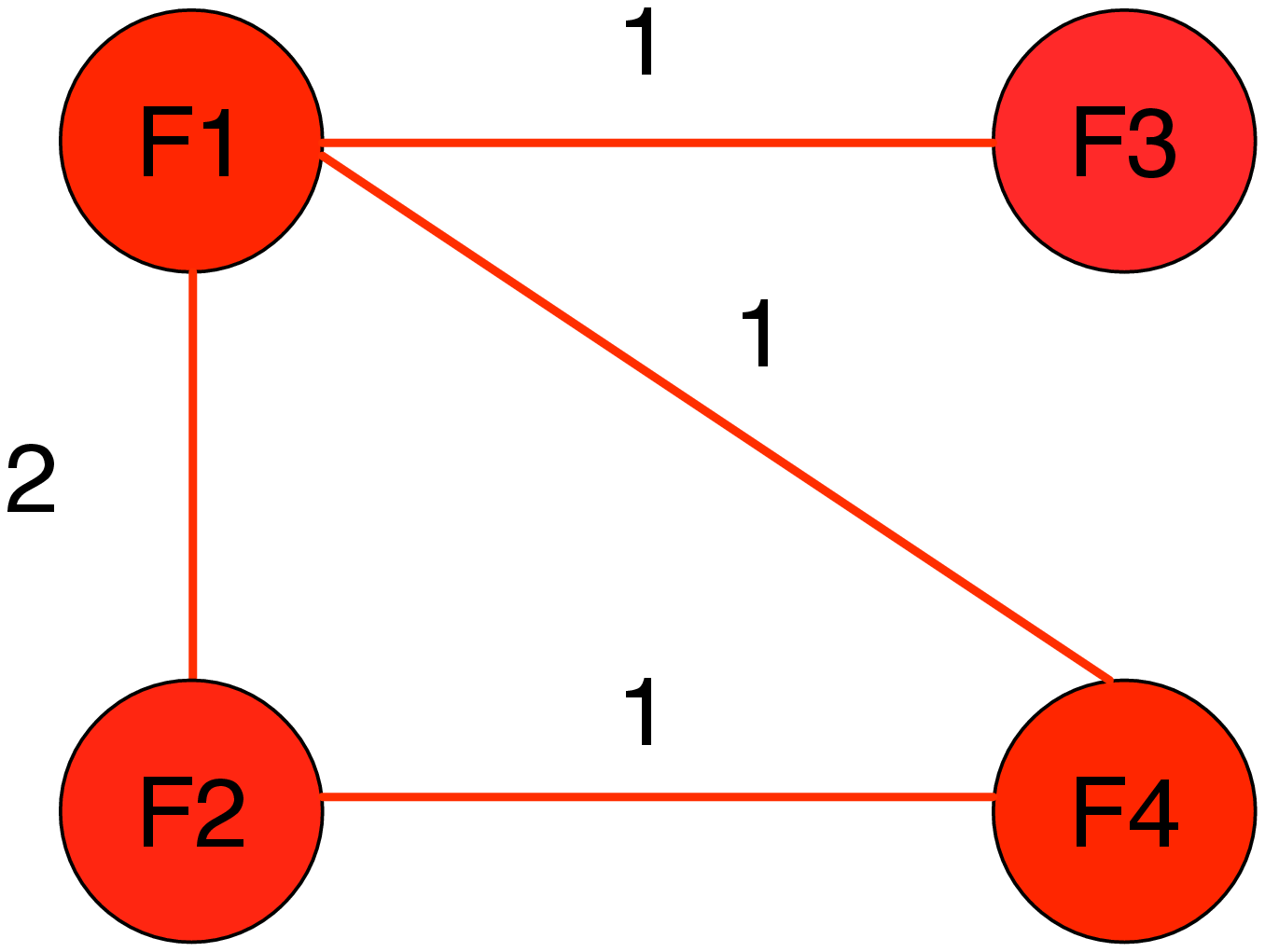}}\hfill
\caption{(a) male sending projection network and (b) female receiving projection network corresponding to the example shown in Figure \ref{fig:example}.}
\label{fig:exampleProjection}
\end{figure}

\subsection{Projection Network Characteristics}

Based on user communication traces, we construct several projection networks for each gender and direction of communications (sending or receiving). The sending projection network is constructed by adding an edge between  
two users who have sent messages to at least one common receiver, while the receiving projection network is 
constructed by add an edge between two users who received messages from at least one common sender. The weight
of each edge in a sending or receiving projection network denotes the number of common receivers or senders 
between the two nodes, respectively. Figure \ref{fig:exampleProjection} illustrates the sending projection network 
of male users and receiving projection network of female users corresponding to the example shown in Figure \ref{fig:example}.

\begin{table}
	\caption{Number of nodes and edges in projection networks.}
	\centering
	\scalebox{0.8}{
		\begin{tabular}{c|c|c|c}
			\hline
			\hline
			
			Network Type & Gender & Nodes & Edges\\
			\hline
			\hline
			\multirow{2}{*}{Sending Projection Network} 
			& Male to Female & 75,379 & 7,716,078 \\
			\cline{2-4}
			& Female to Male & 28,550 & 1,025,738\\
			\hline
			\multirow{2}{*}{Receiving Projection Network} 
			& Male from Female & 43,420 & 22,603,491 \\
			\cline{2-4}
			& Female from Male & 45,214 & 18,858,211\\
			\hline
			\hline
		\end{tabular}
	}
	\label{table: network_attributes}
\end{table}

Table \ref{table: network_attributes} describes several important network measurements of the sending and receiving projection networks for male and female users.

\begin{figure}
\subfloat[\label{fig:sendDegree}]
  {\includegraphics[width=.5\linewidth]{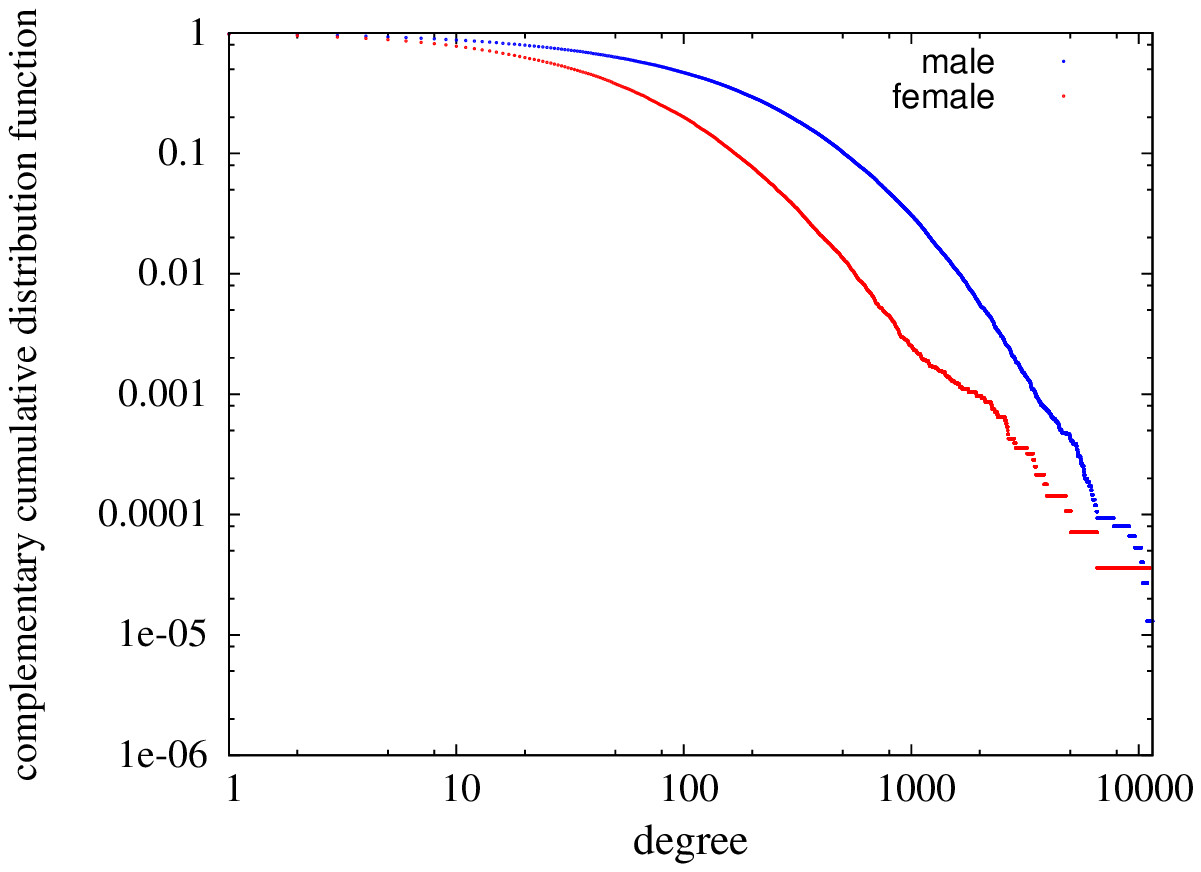}}\hfill
\subfloat[\label{fig:sendWeight}]
  {\includegraphics[width=.5\linewidth]{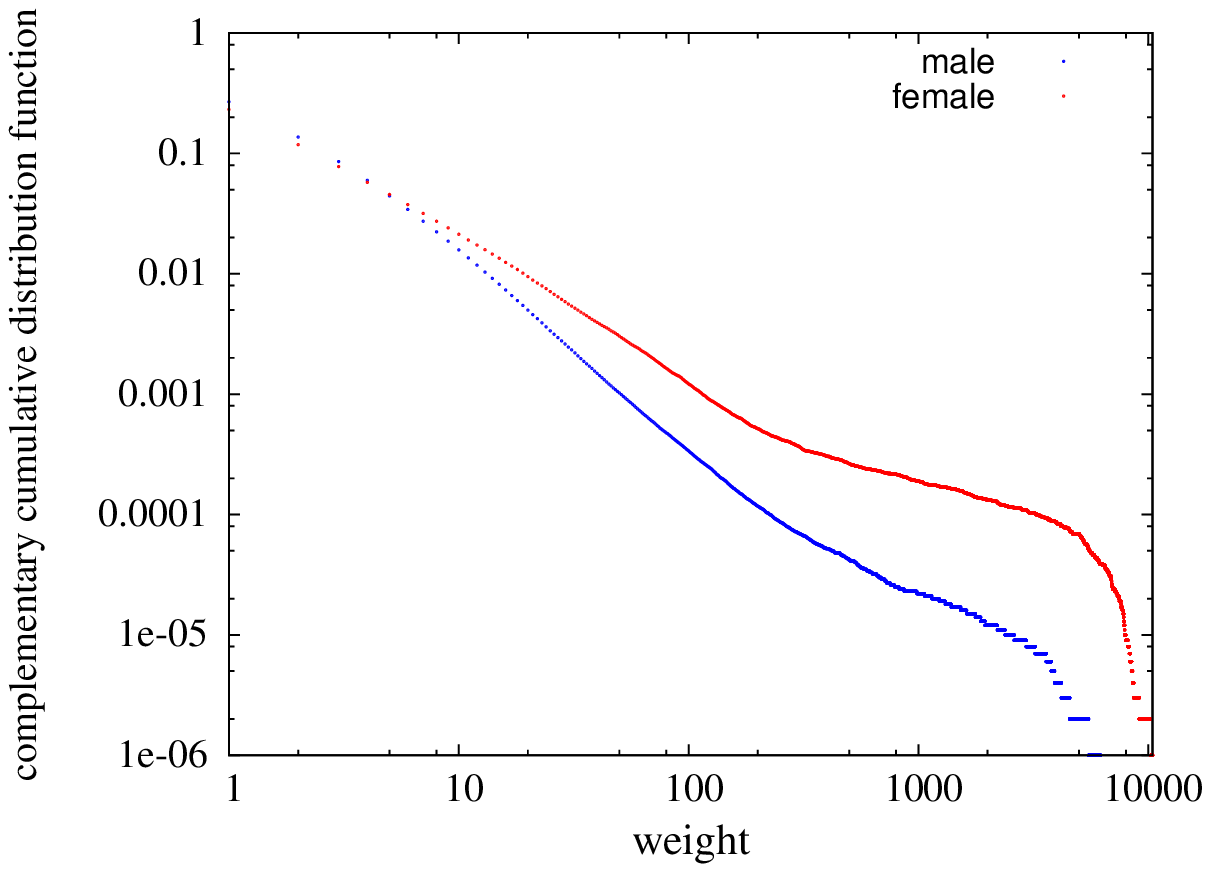}}\hfill
\caption{CCDF for (a) degree and (b) weight distribution for the sending projection networks.}
\label{fig:sendProjection}
\end{figure}

Figure \ref{fig:sendProjection} shows the node degree and edge weight distributions for the sending projection network of both male and female users. The node degree in a sending projection network represents the number of other users with whom the user share some degree of similar interest. We observe that a male shares similar interest with a larger number of peers than a female. The median degrees for male and female users are 89 and 33, respectively. We also observe that most edges in the sending projection networks have a low weight (i.e., most pairs of users contact very few common receivers), in particular, 73.1\% and 76.8\% of the edges in the male and female sending projection networks have a weight of 1. Also, the CCDF of edge weight distribution of females lies above that of males, indicating that females tend to share more common receivers than males.

\begin{figure}
\subfloat[\label{fig:receiveDegree}]
  {\includegraphics[width=.5\linewidth]{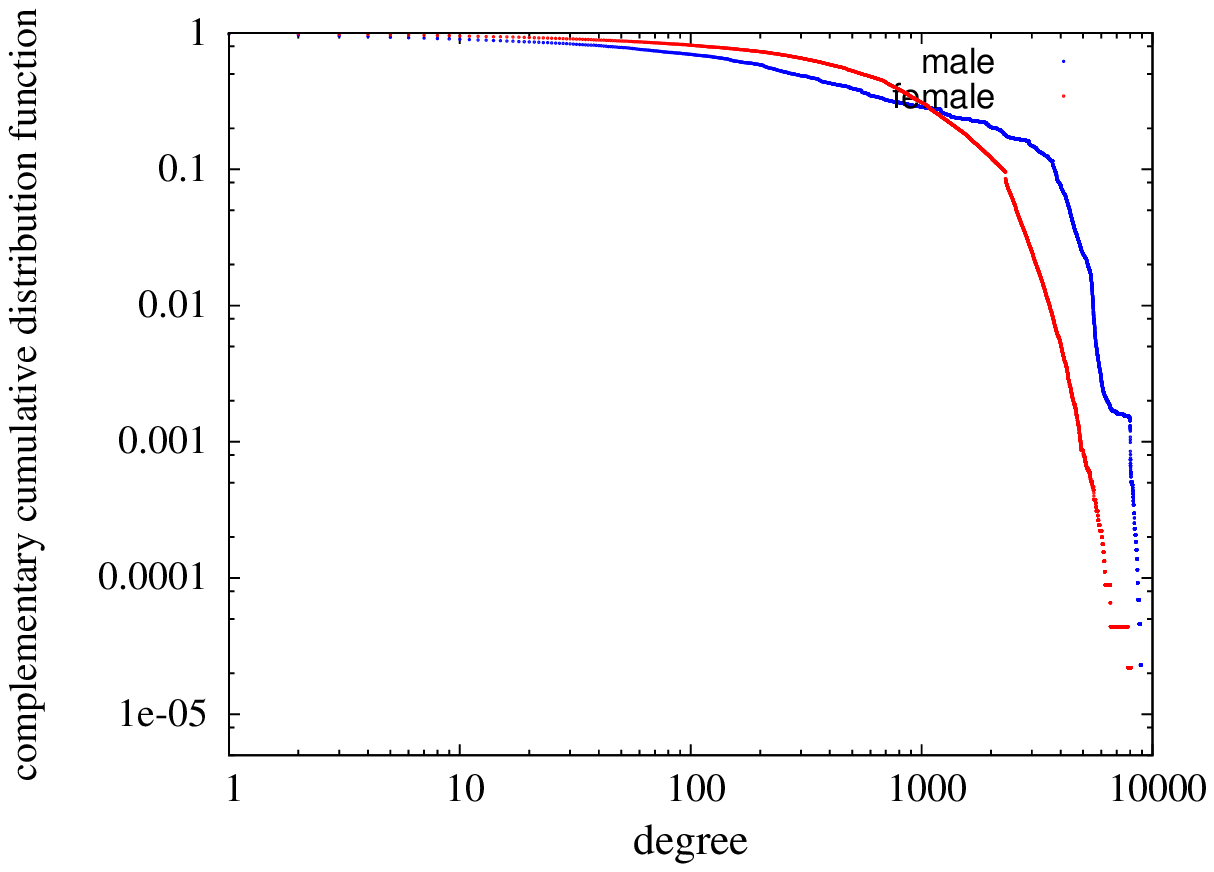}}\hfill
\subfloat[\label{fig:receiveWeight}]
  {\includegraphics[width=.5\linewidth]{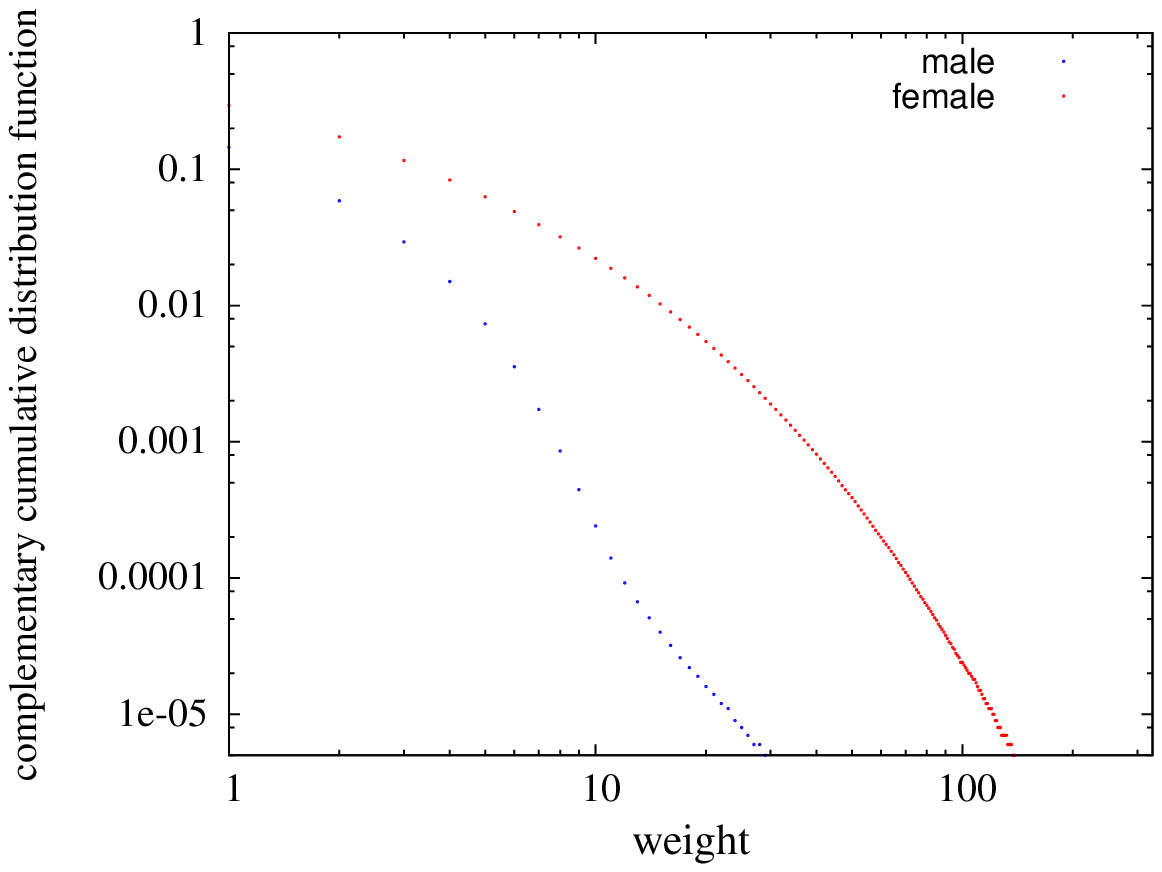}}\hfill
\caption{CCDF for (a) degree and (b) weight distribution for the receiving projection networks.}
\label{fig:receiveProjection}
\end{figure}

Figure \ref{fig:receiveProjection} plots the distributions of node degree and edge weight of the receiving projection network for both male and female users. The median degrees for male and female users are 283 and 551, respectively. Most of the edges have a low weight (i.e., most pairs of users are contacted by very few common senders), in particular, 
85.5\% and 70.6\% of the edges in the male and female receiving projection networks have a weight of 1.

\section{Evaluation} \label{sec:evaluation}
For a given service users in our test set, we rank the recommended users by comparing their reciprocal 
scores, and recommend the top-K users in the list. We evaluate the performance of each algorithm by comparing 
the top-K users in the recommended list with the receivers contacted by the service user in test set.

\subsection{Experiment Setup}

Most of the active users in our dataset are newly registered users. They are usually very active in looking for 
a potential date in the first one or two weeks after registration \cite{Peng2013}. We select the user interactions 
within 10 days from user registration time as the training data, and interactions in the remaining 
time as the test set.

We filter the service users by selecting users who have sent or replied at least 5 
messages in the training period, and use the interactions between these users in the test period as the 
test set. For the training set, we count all the interactions initiated, received or replied to by the selected 
users in the training period, and use these interactions to train our recommendation system. 
Table \ref{table:experiment} summarizes the experiment dataset.


\begin{table}
\caption{Experiment Dataset}
\centering
\scalebox{1}{
  \begin{tabular}{|c|c| p{2.1cm} | p{2.1cm} |}
  \hline
  \hline
  Male & Female & \# of Messages in Training Set & \# of Messages in Test Set\\
  \hline
  24,602 & 8,250 & 730,110 & 270,294  \\
  \hline
  \end{tabular}
}
\label{table:experiment}
\end{table}

For RECON (CB1) and CB2,  we manually pick 20 features over all 39 features. These selected features include age, height, weight, city, education level, income, house status, marriage status, children 
status, physical looking, car status, number of photos, smoking habit, drinking habit, marriage status, parents status, children plan, dating method, and wedding plan. Among these features, age, height, weight, and number of photos are treated as numeric values. For HCF, we performed several experiments to get the optimal weight parameter $s$ in the computation of the success score between two users \cite{Kang2013}.

\subsection{Evaluation Metrics}

For a given service user, we define three set of users: $T$ as the set
of users we have recommended to the service user, $I$ as the collections of users who have been contacted by 
the service user and $R$ as the set of users who have been contacted by the service user and replied to the 
service user in the test set. We define the following two different evaluation metrics:
\begin{equation}
	\begin{aligned}
	I\text{-}Precision = \frac{|I \cap T|}{|T|}, \;
	I\text{-}Recall = \frac{|I \cap T|}{|I|},
	\end{aligned}
	\label{formula:IC}
\end{equation}
and 
\begin{equation}
	\begin{aligned}
	R\text{-}Precision = \frac{|R \cap T|}{|T|}, \;
	R\text{-}Recall = \frac{|R \cap T|}{|R|},
	\end{aligned}
	\label{formula:RC}
\end{equation}
where $I\text{-}Precision$ and $R\text{-}Precision$ measure the ratio of users in the recommendation list who have been contacted by or exchanged messages with the service user, respectively. $I\text{-}Recall$ and $R\text{-}Recall$ measure  the ratio of users who have been contacted by or exchanged messages with the service user in the list of recommended users. From another perspective, $I\text{-}Precision$ and $I\text{-}Recall$ measure an algorithm's performance in recommending users that the service user is interested in and thus likely to contact, and $R\text{-}Precision$ and $R\text{-}Recall$ measure an algorithm's performance in recommending users who have mutual interest with the service user and are thus likely to reciprocate when contacted. 

\begin{figure}
\subfloat[\label{fig:MaCBIP}]
  {\includegraphics[width=.5\linewidth]{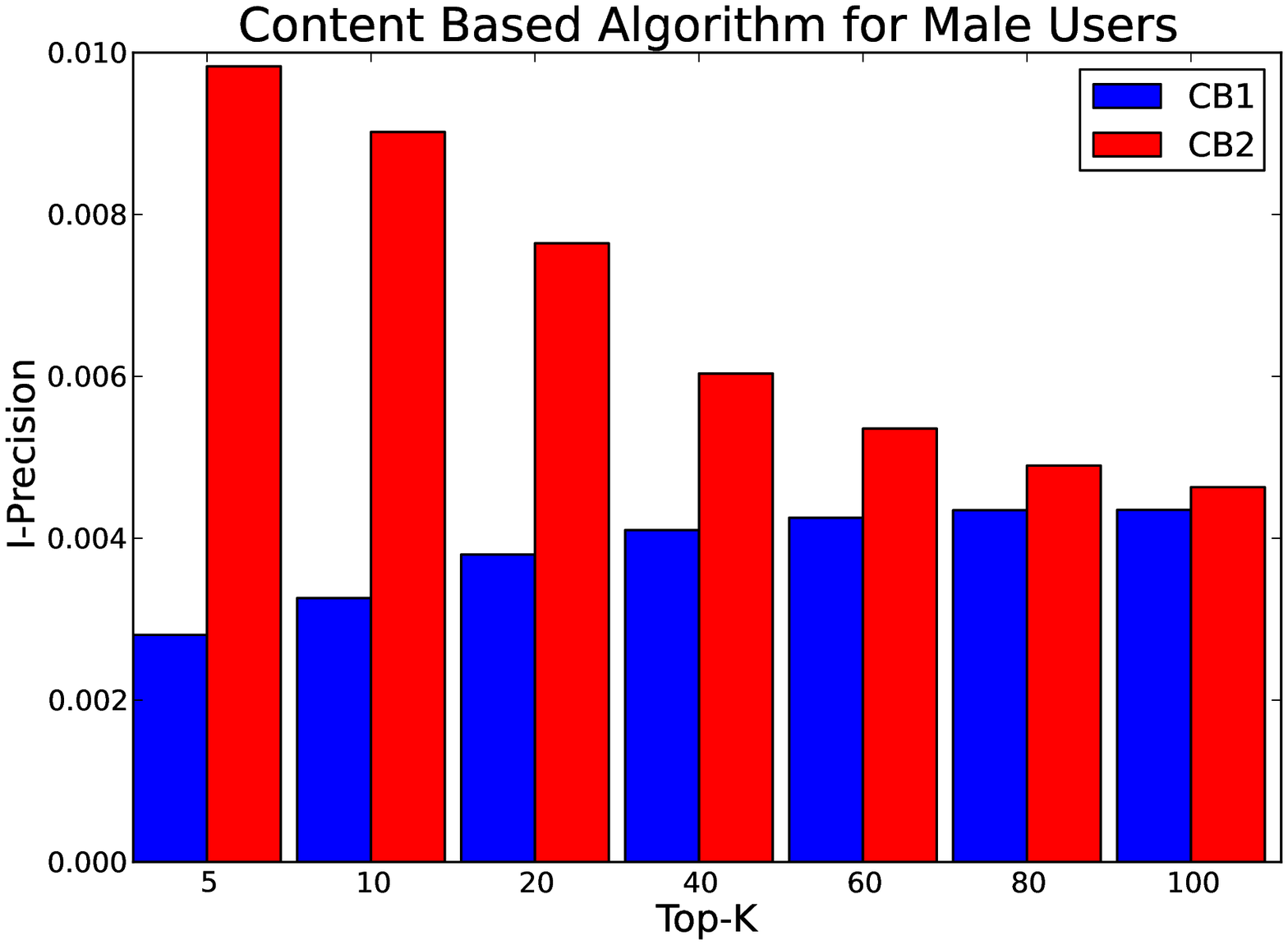}}\hfill
\subfloat[\label{fig:MaCBIR}]
  {\includegraphics[width=.5\linewidth]{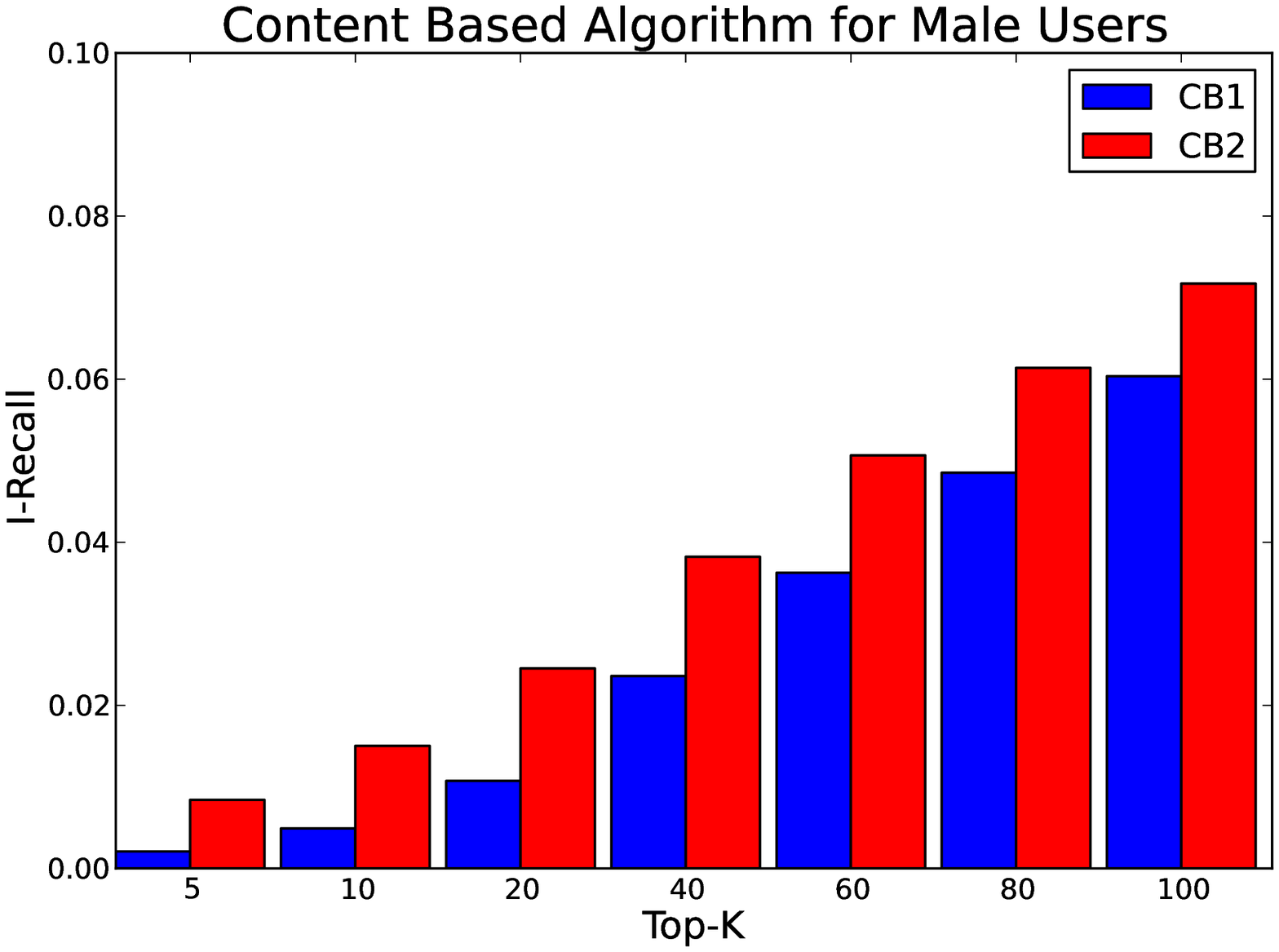}}\hfill
\caption{$I\text{-}Precision$ and $I\text{-}Recall$ of content-based algorithms for male users}
\label{fig:content_i_m}
\end{figure}

\begin{figure}
\subfloat[\label{fig:FCBIP}]
  {\includegraphics[width=.5\linewidth]{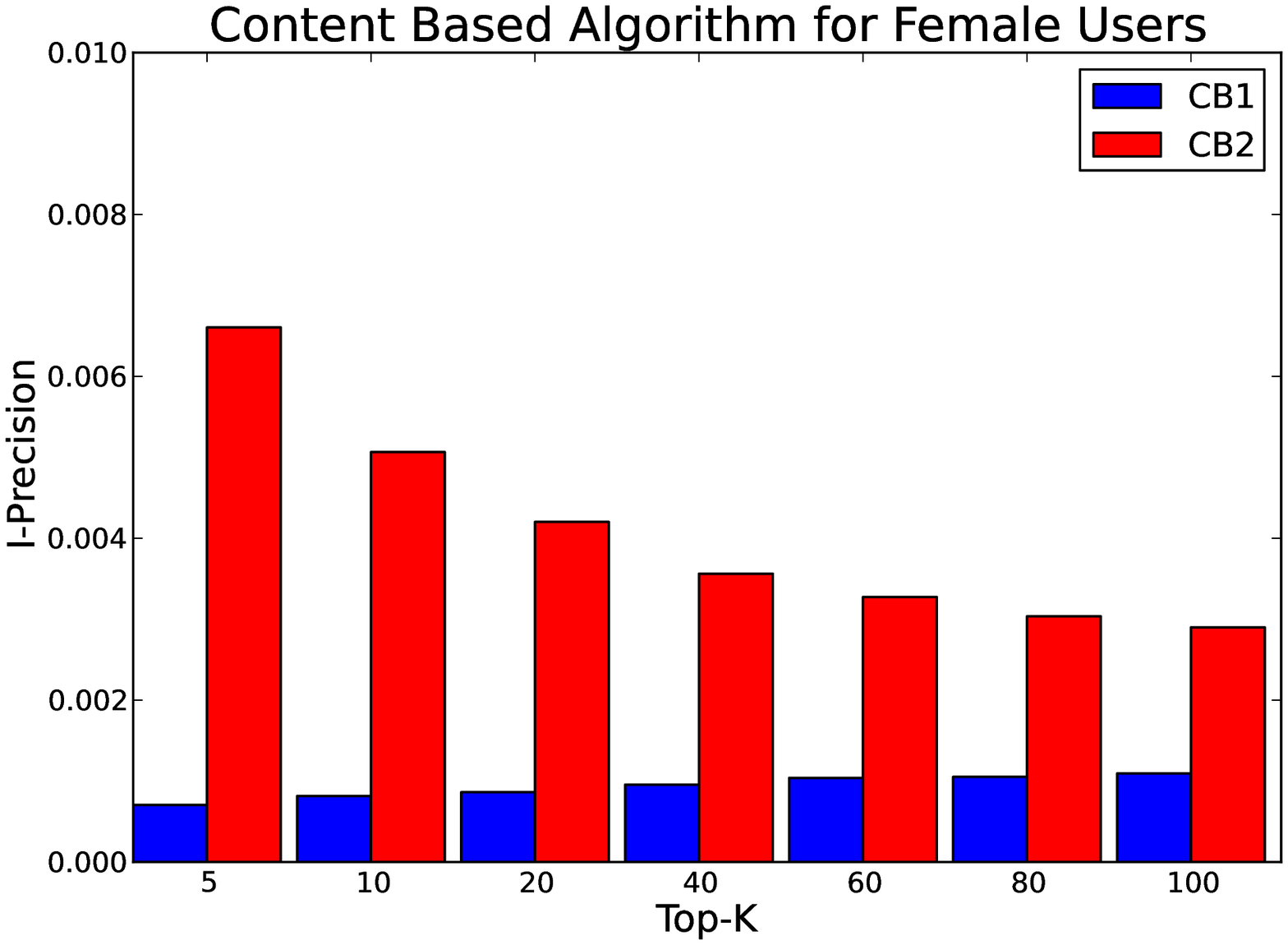}}\hfill
\subfloat[\label{fig:FCBIR}]
  {\includegraphics[width=.5\linewidth]{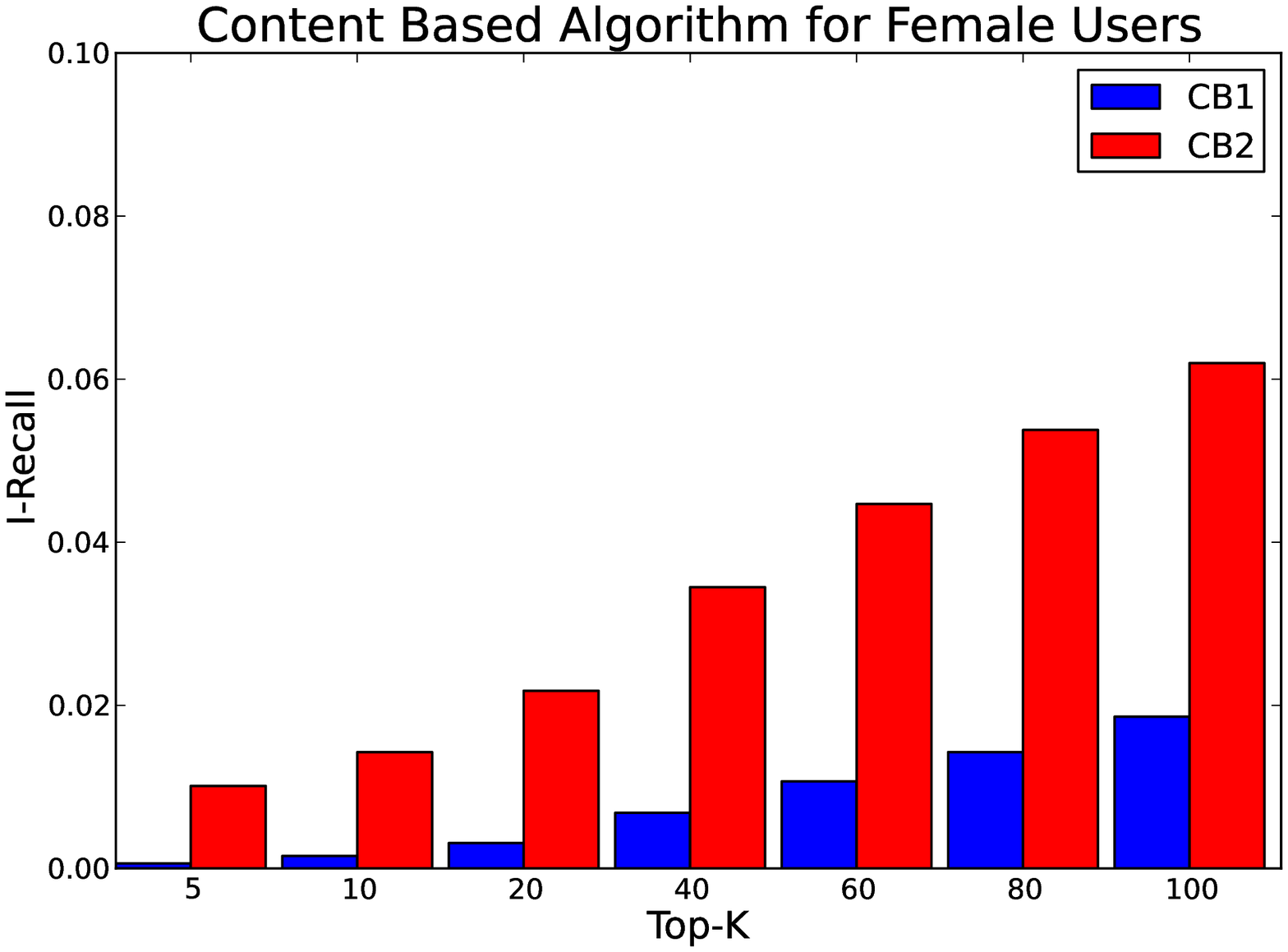}}\hfill
\caption{$I\text{-}Precision$ and $I\text{-}Recall$ of content-based algorithms for female users}
\label{fig:content_i_f}
\end{figure}

\subsection{Evaluation Results}
In this subsection, we apply our recommendation algorithms on the experiment datasets, and compare the performance with RECON \cite{Luiz2010} and HCF \cite{Kang2013}. We first report the performance of recommending users that the service user will contact, and then evaluate the performance of recommending users who will reciprocate when contacted by the service user.




\begin{figure}
\subfloat[\label{fig:MaCFIP}]
  {\includegraphics[width=.5\linewidth]{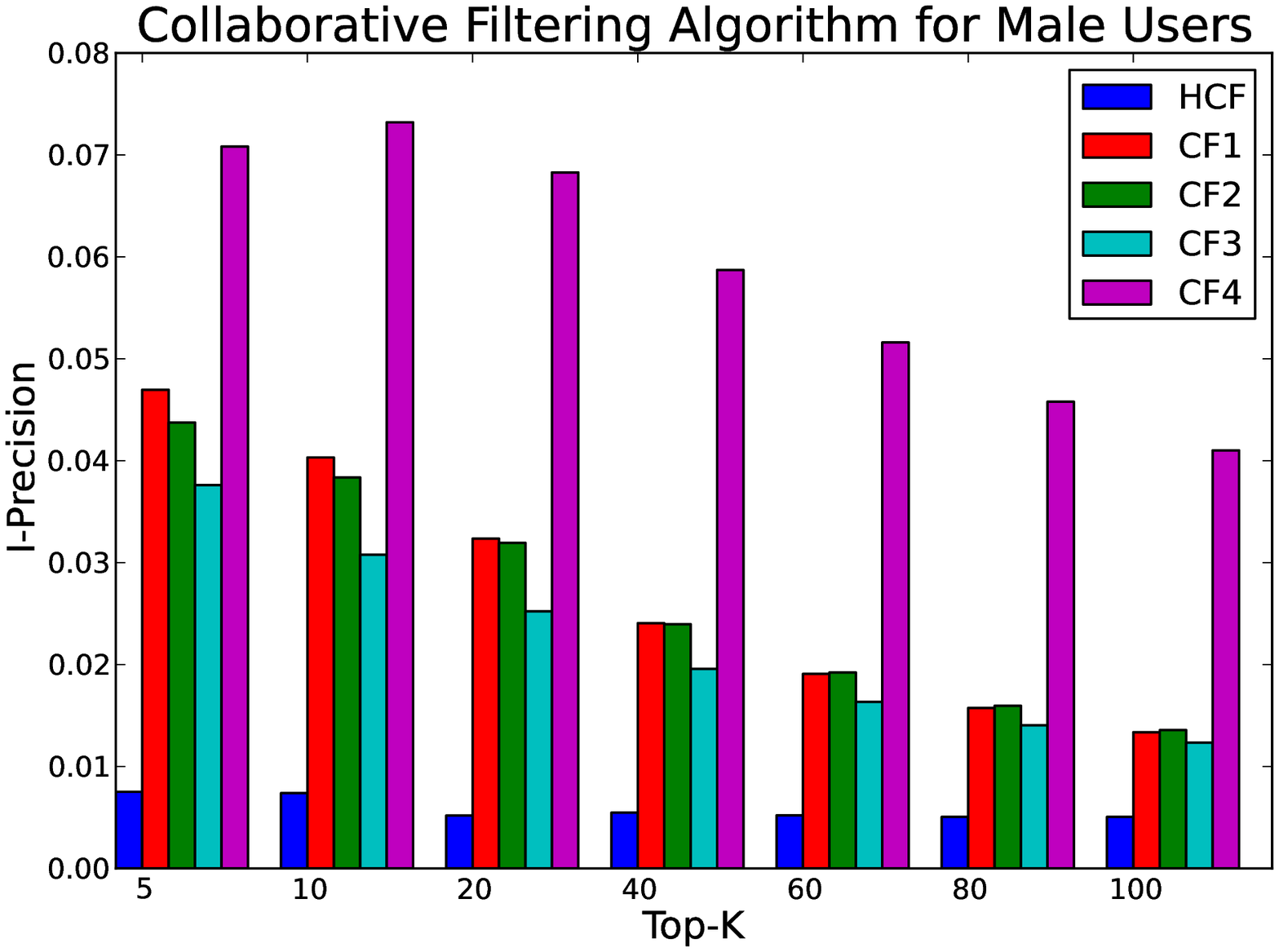}}\hfill
\subfloat[\label{fig:MaCFIR}]
  {\includegraphics[width=.5\linewidth]{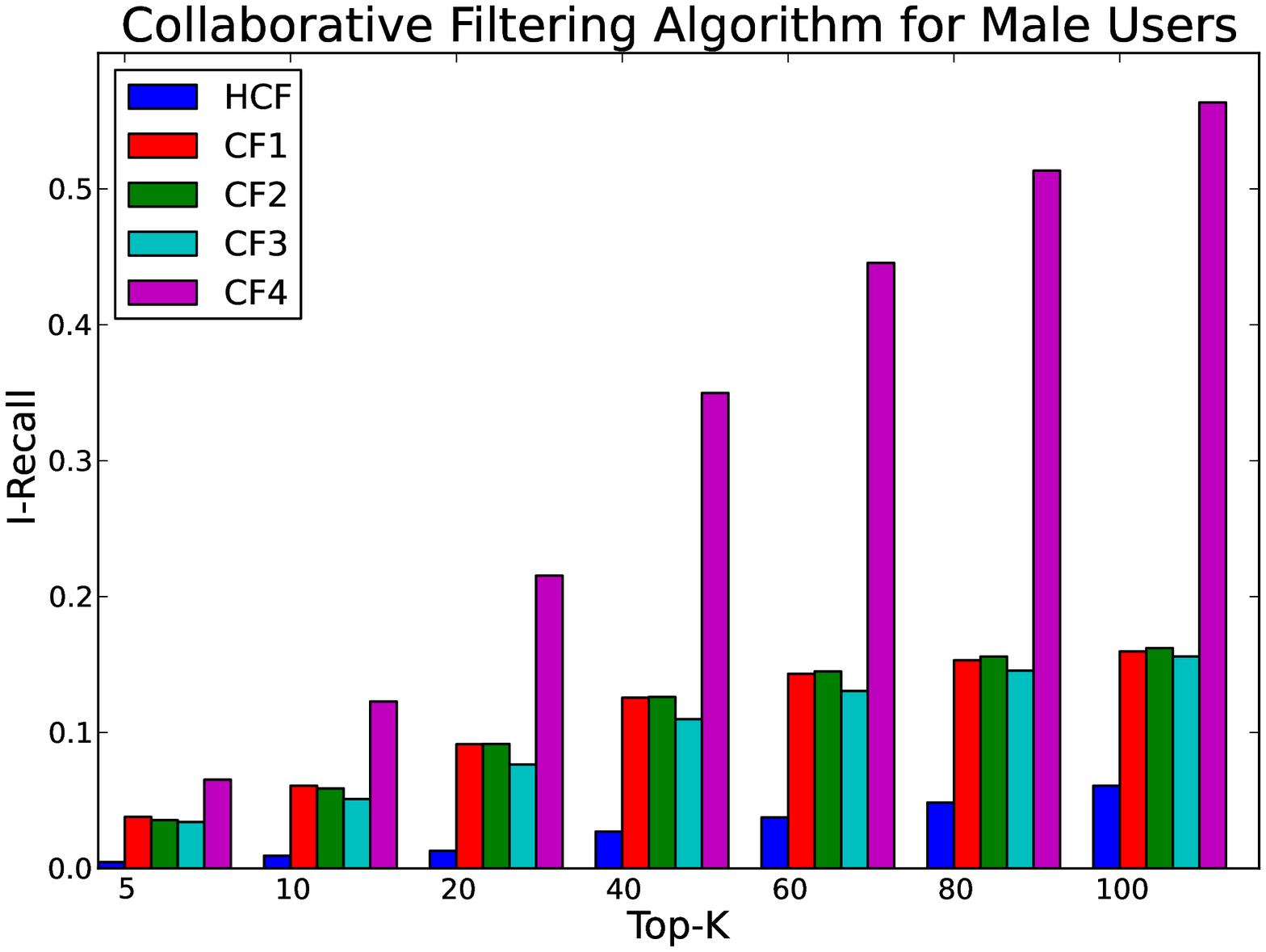}}\hfill
\caption{$I\text{-}Precision$ and $I\text{-}Recall$ of collaborative filtering algorithms for male users}
\label{fig:cf_i_m}
\end{figure}

\begin{figure}
\subfloat[\label{fig:FCFIP}]
  {\includegraphics[width=.5\linewidth]{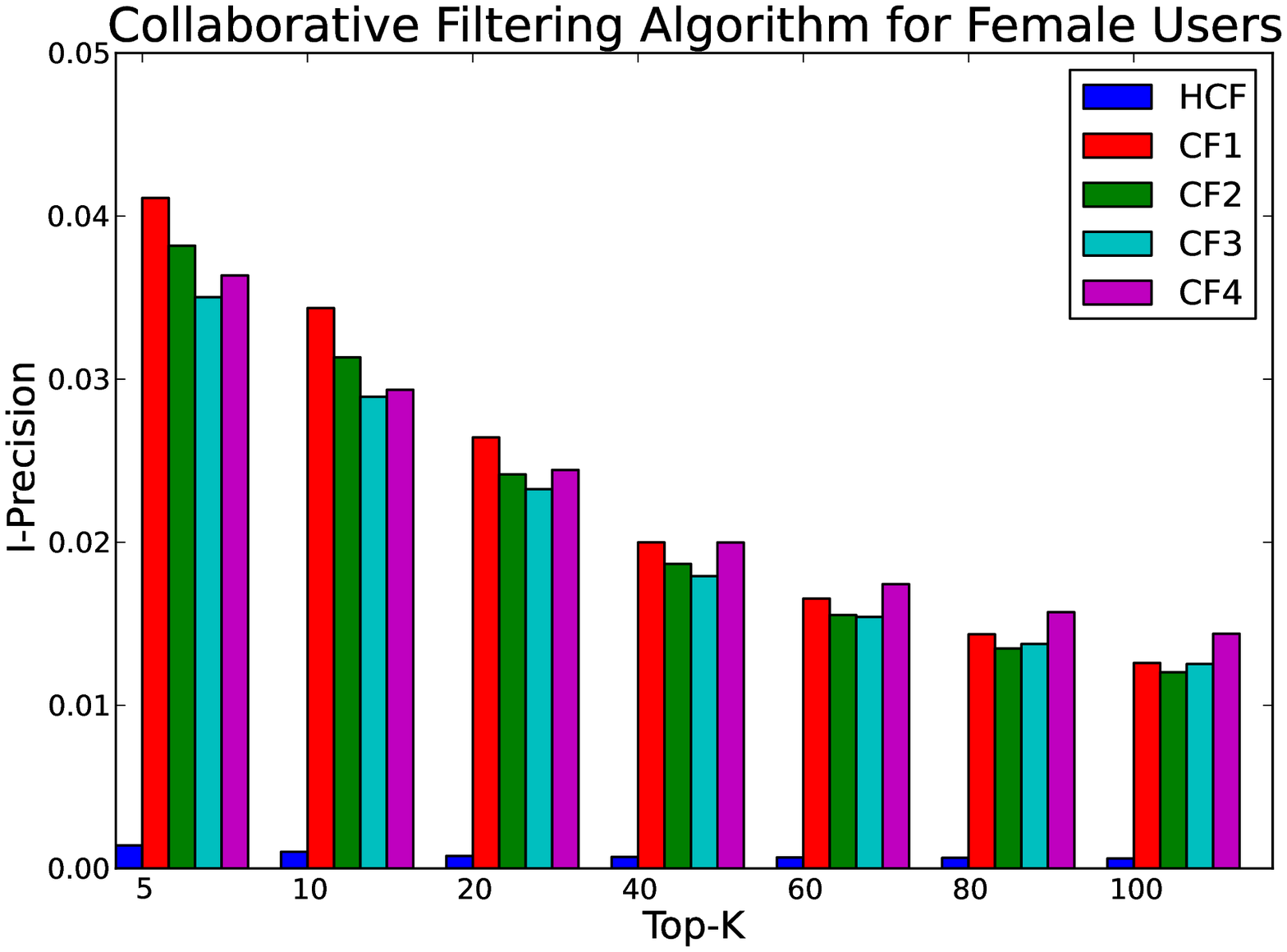}}\hfill
\subfloat[\label{fig:FCFIR}]
  {\includegraphics[width=.5\linewidth]{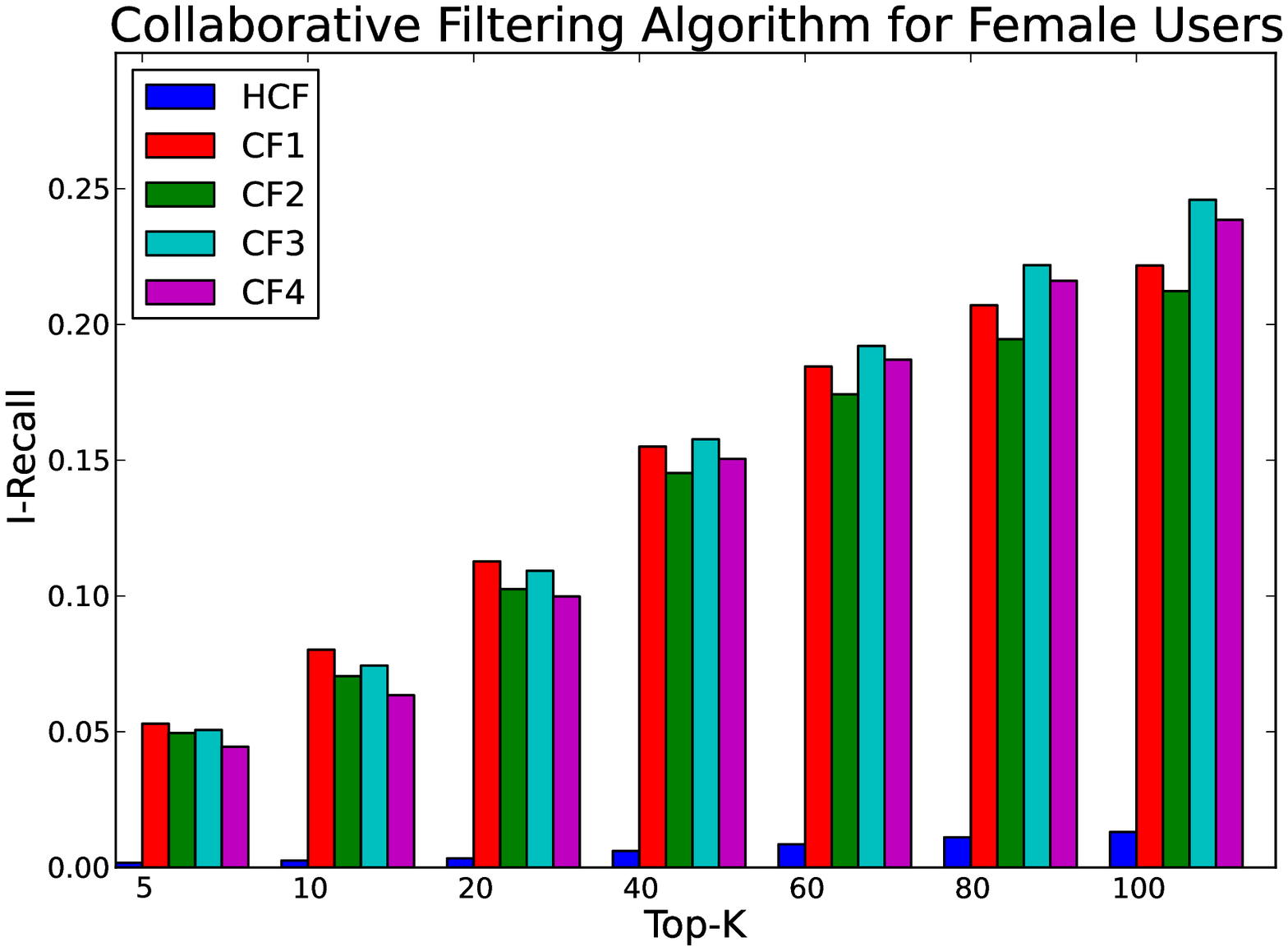}}\hfill
\caption{$I\text{-}Precision$ and $I\text{-}Recall$ of collaborative filtering algorithms for female users}
\label{fig:cf_i_f}
\end{figure}

\subsubsection{$I\text{-}Precision$ and $I\text{-}Recall$}
We first examine the performance of these algorithms in recommending users whom the service user is interested in and thus likely to contact.

Figures \ref{fig:content_i_m} and \ref{fig:content_i_f} show the performance of the two content-based algorithms, namely, CB1(RECON) and CB2. We observe that by preserving the values of numeric attributes in the similarity measure, CB2 significantly outperforms CB1 in both precision and recall. The improvement is more pronounced for females than for males.

\begin{figure}
\subfloat[\label{fig:MCBRP}]
  {\includegraphics[width=.5\linewidth]{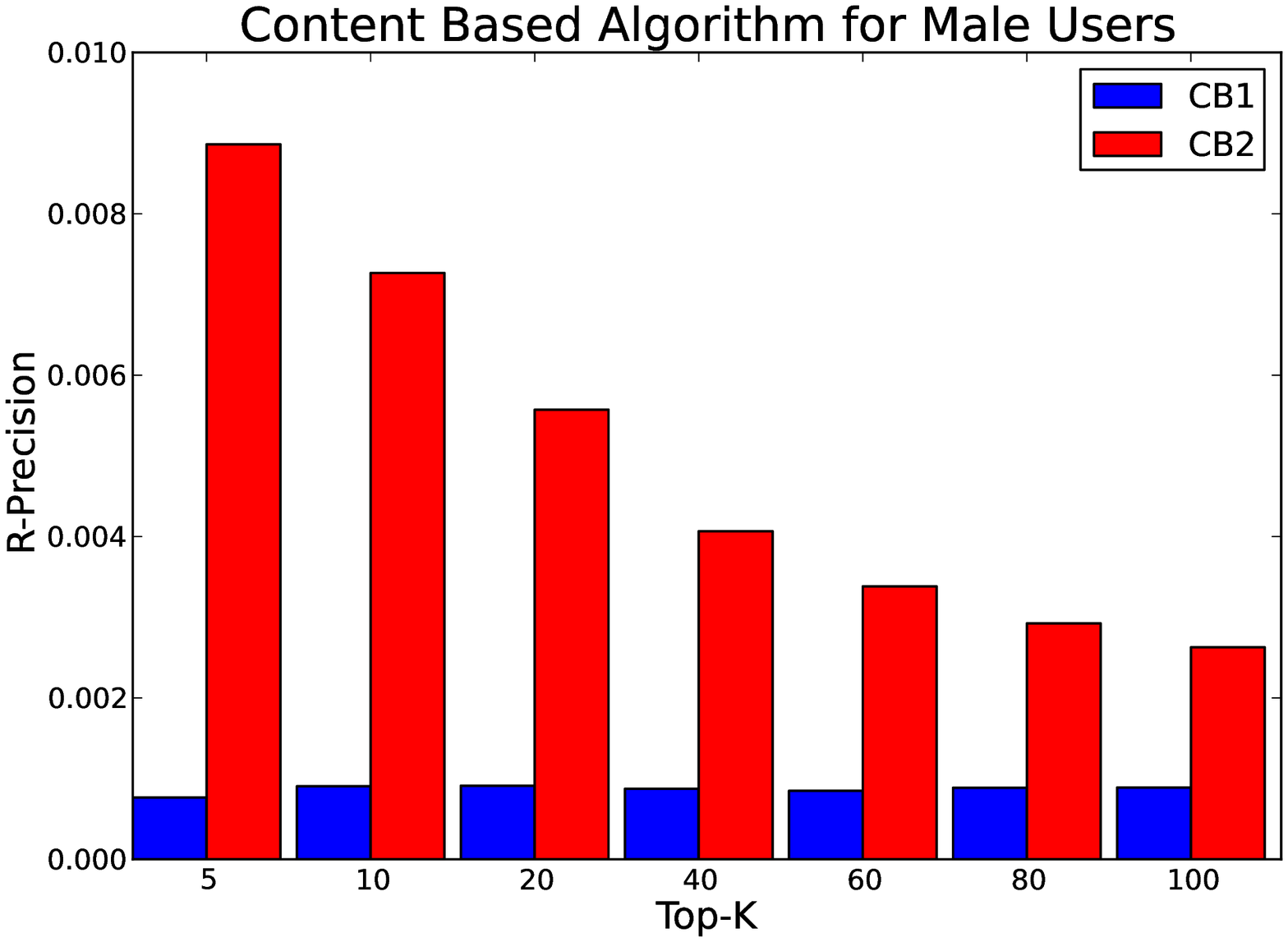}}\hfill
\subfloat[\label{fig:MCBRR}]
  {\includegraphics[width=.5\linewidth]{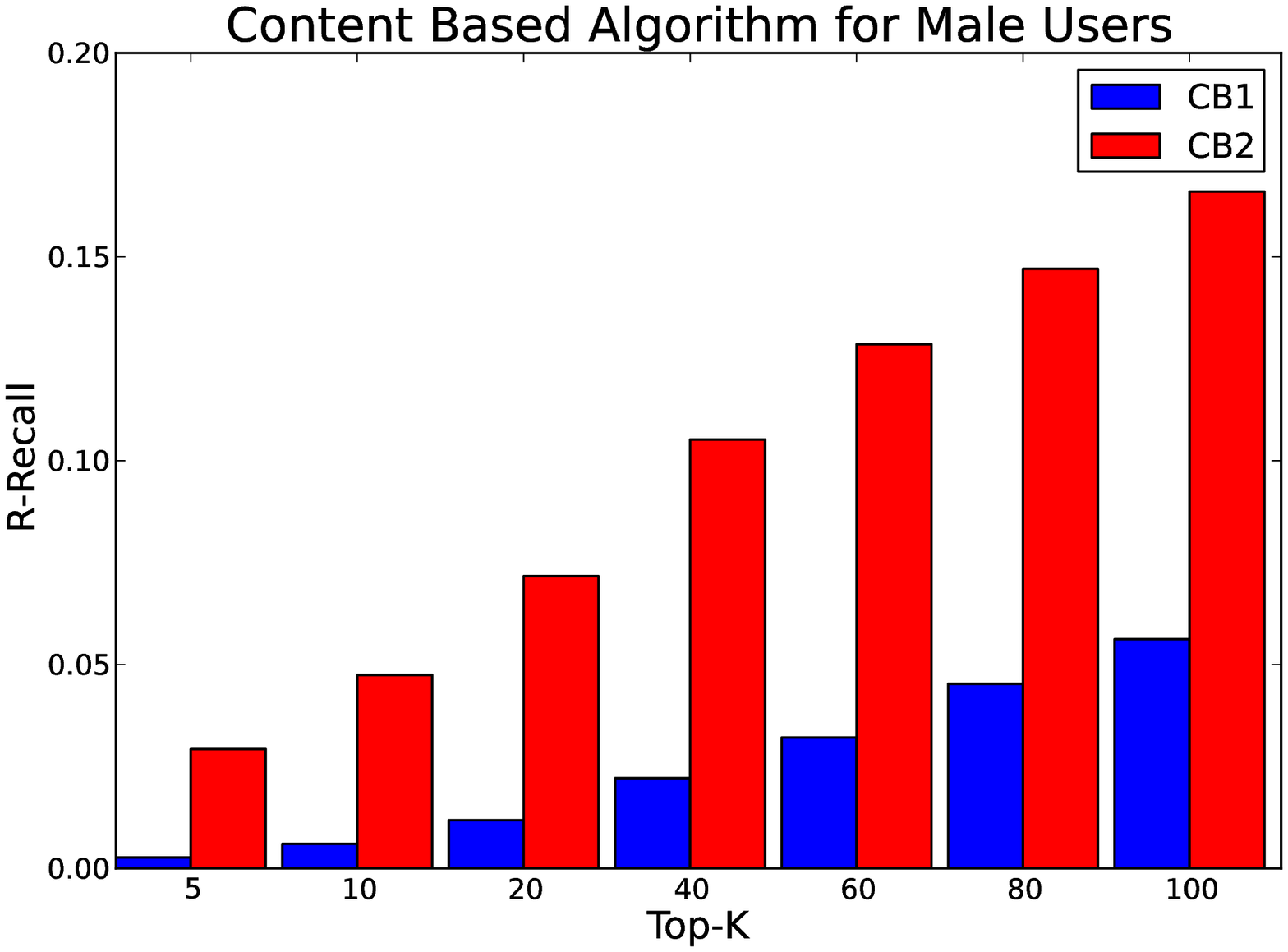}}\hfill
\caption{$R\text{-}Precision$ and $R\text{-}Recall$ of content-based algorithms for male users.}
\label{fig:content_r_m}
\end{figure}

Figures \ref{fig:cf_i_m} and \ref{fig:cf_i_f} show the $I\text{-}Precision$ and $I\text{-}Recall$ of the collaborative filtering based recommendation algorithms for male and female users, respectively. We observe that for both male and female users, the four algorithms proposed in this paper, CF1 to CF4, significantly outperform previously proposed HCF algorithm. For male users, while there are some difference in the performance of CF1, CF2, and CF3, the difference is rather small when compared with CF4, which is much more effective in attracting the service user to contact the recommended users. However, for female users, the four algorithms (CF1 to CF4) all perform similarly. There is no algorithm significantly outperforming others. Note that the CF4 algorithm captures the interest of the service user in recommended users as well as the attractiveness of the recommended users to the service user. The results indicate that when it comes to looking for potential dates, males tend to be more focused on their own interest and oblivious towards their attractiveness to potential dates, while females do not show such behavior.

\begin{figure}
\subfloat[\label{fig:FCBRP}]
  {\includegraphics[width=.5\linewidth]{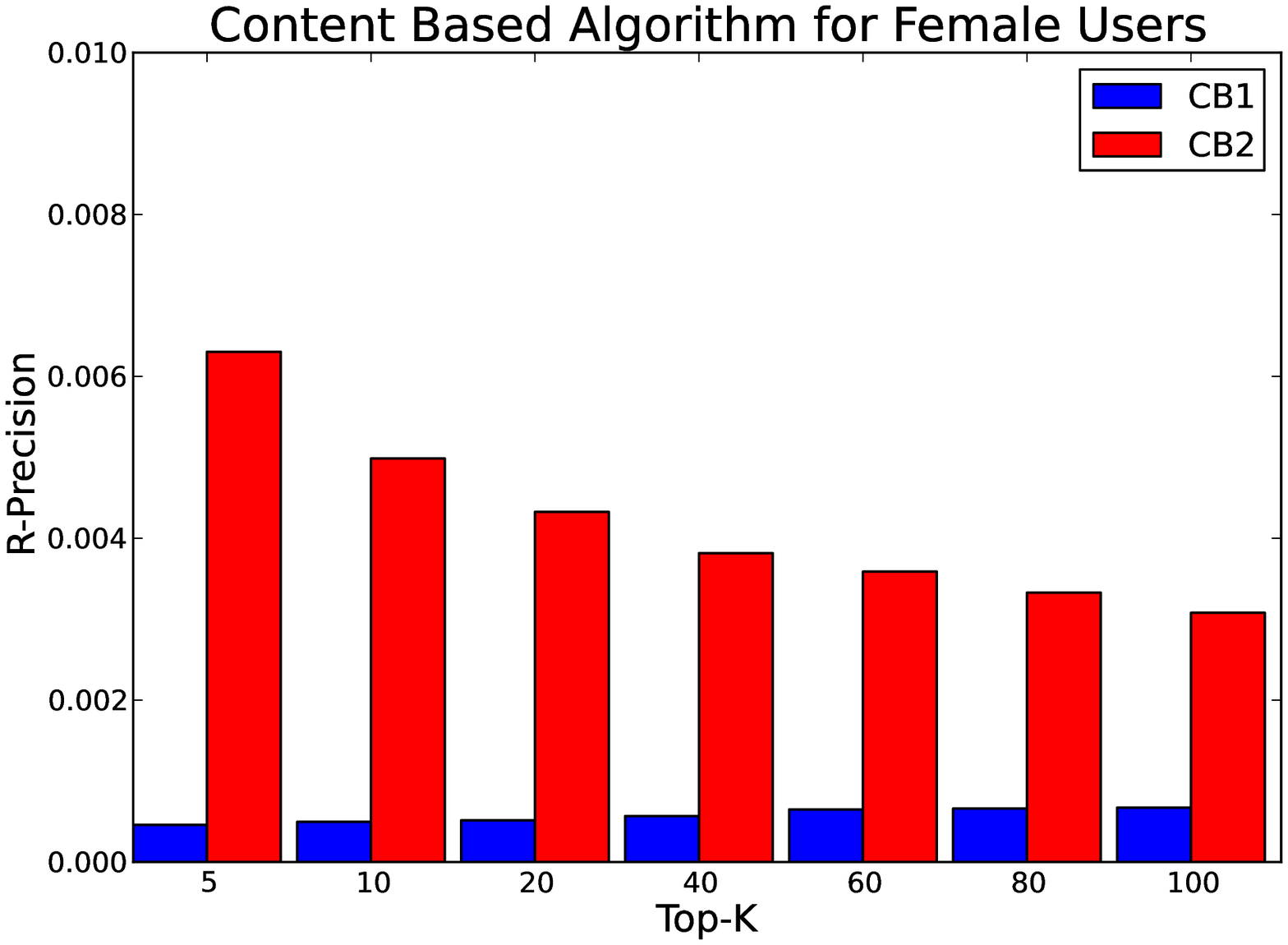}}\hfill
\subfloat[\label{fig:FCBRR}]
  {\includegraphics[width=.5\linewidth]{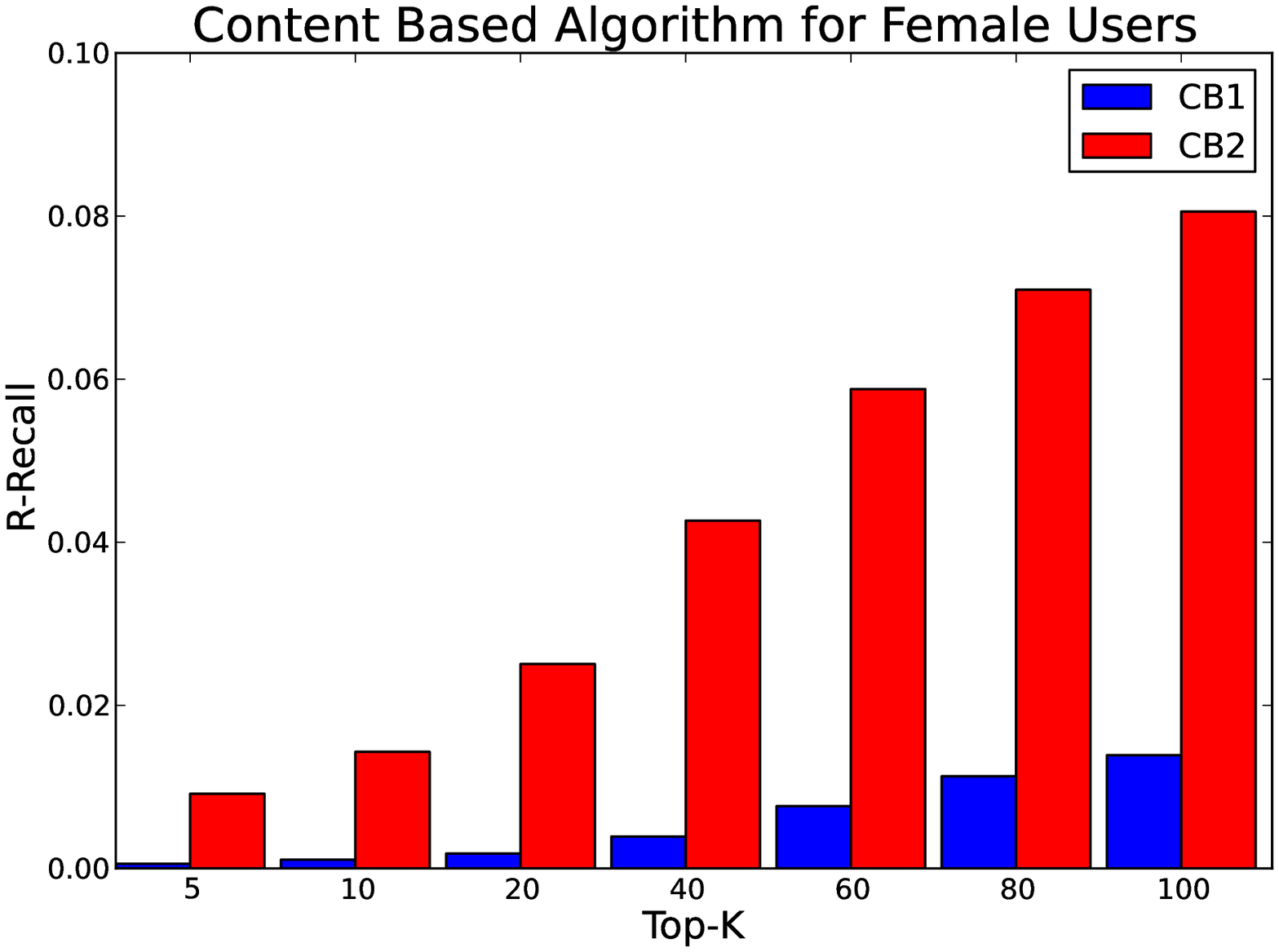}}\hfill
\caption{$R\text{-}Precision$ and $R\text{-}Recall$ of content-based algorithms for female users.}
\label{fig:content_r_f}
\end{figure}

\subsubsection{$R\text{-}Precision$ and $R\text{-}Recall$}
We now examine the performance of these algorithms in recommending users who will be contacted by and exchange messages with the service user. 

Figures \ref{fig:content_r_m} and \ref{fig:content_r_f} show the $R\text{-}Precision$ and $R\text{-}Recall$ of the two content-based algorithms for male and female users, respectively. Similar to $I\text{-}Precision$ and $I\text{-}Recall$, CB2 performs much better than CB1(RECON), as CB2 does not convert numeric attributes into nominal attributes and thus does not loss information of these numeric attributes.

\begin{figure}
\subfloat[\label{fig:MaCFRP}]
  {\includegraphics[width=.5\linewidth]{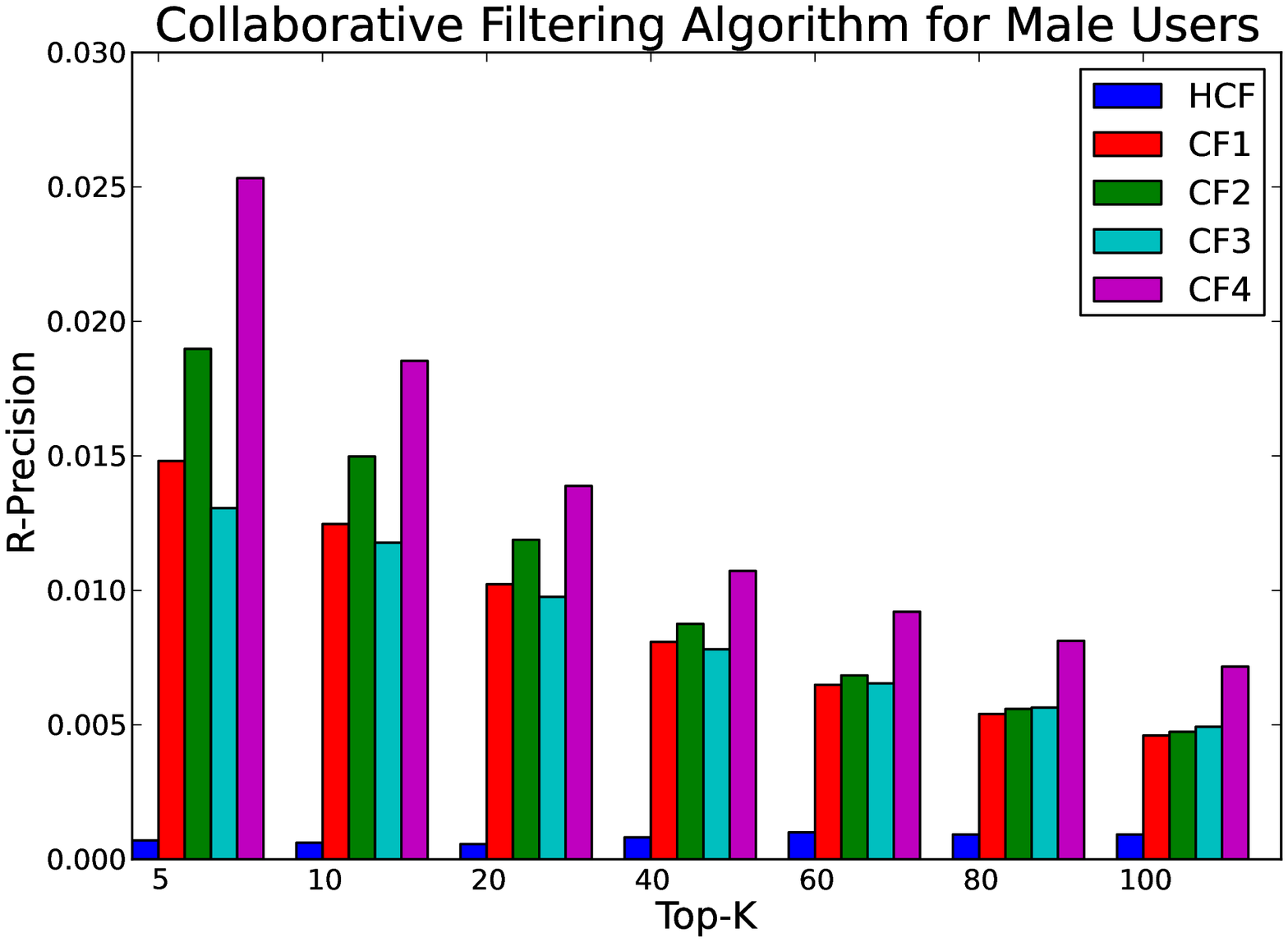}}\hfill
\subfloat[\label{fig:MaCFRR}]
  {\includegraphics[width=.5\linewidth]{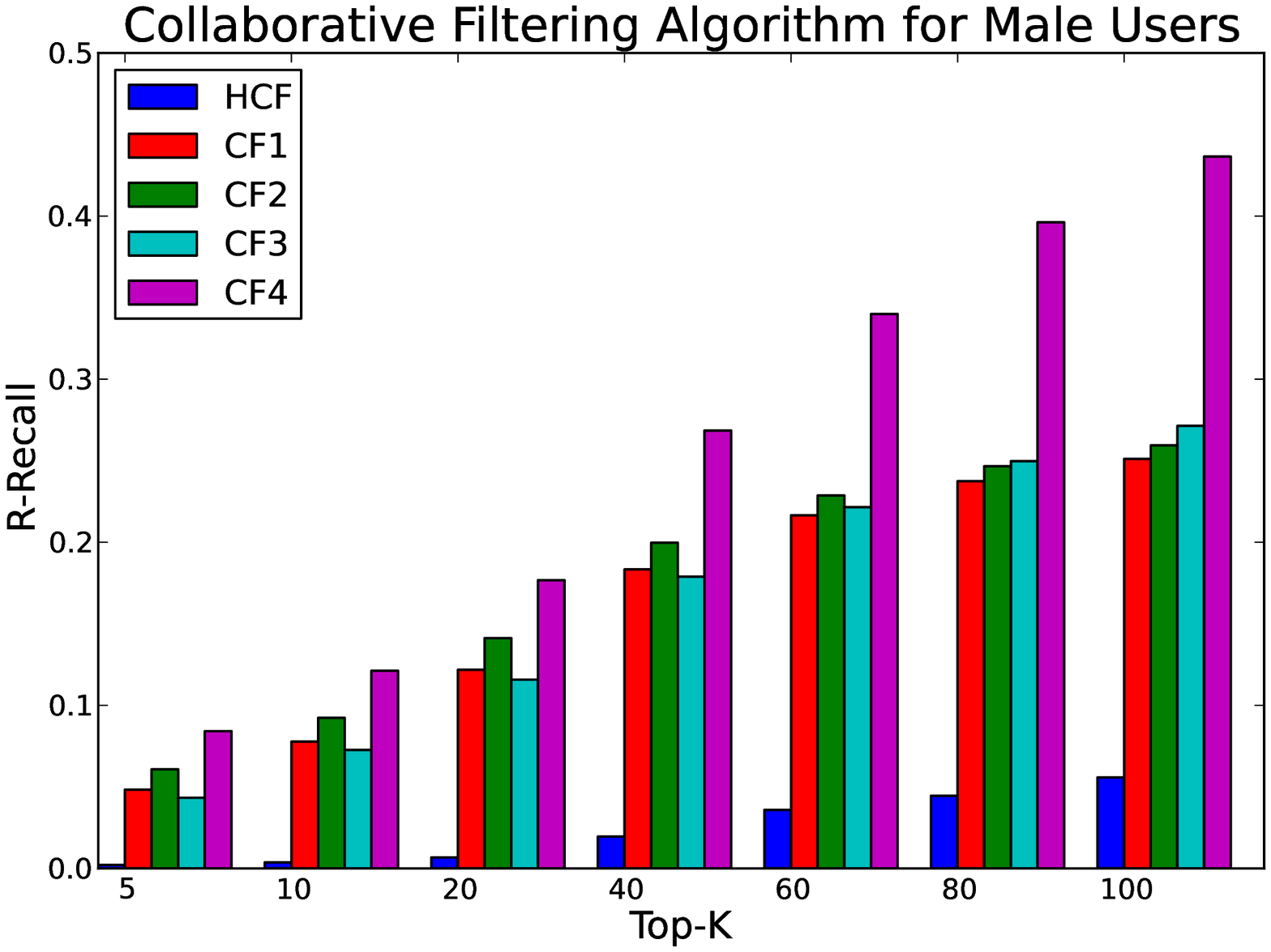}}\hfill
\caption{$R\text{-}Precision$ and $R\text{-}Recall$ of collaborative filtering algorithms for male users.}
\label{fig:cf_r_m}
\end{figure}

\begin{figure}
\subfloat[\label{fig:FCFRP}]
  {\includegraphics[width=.5\linewidth]{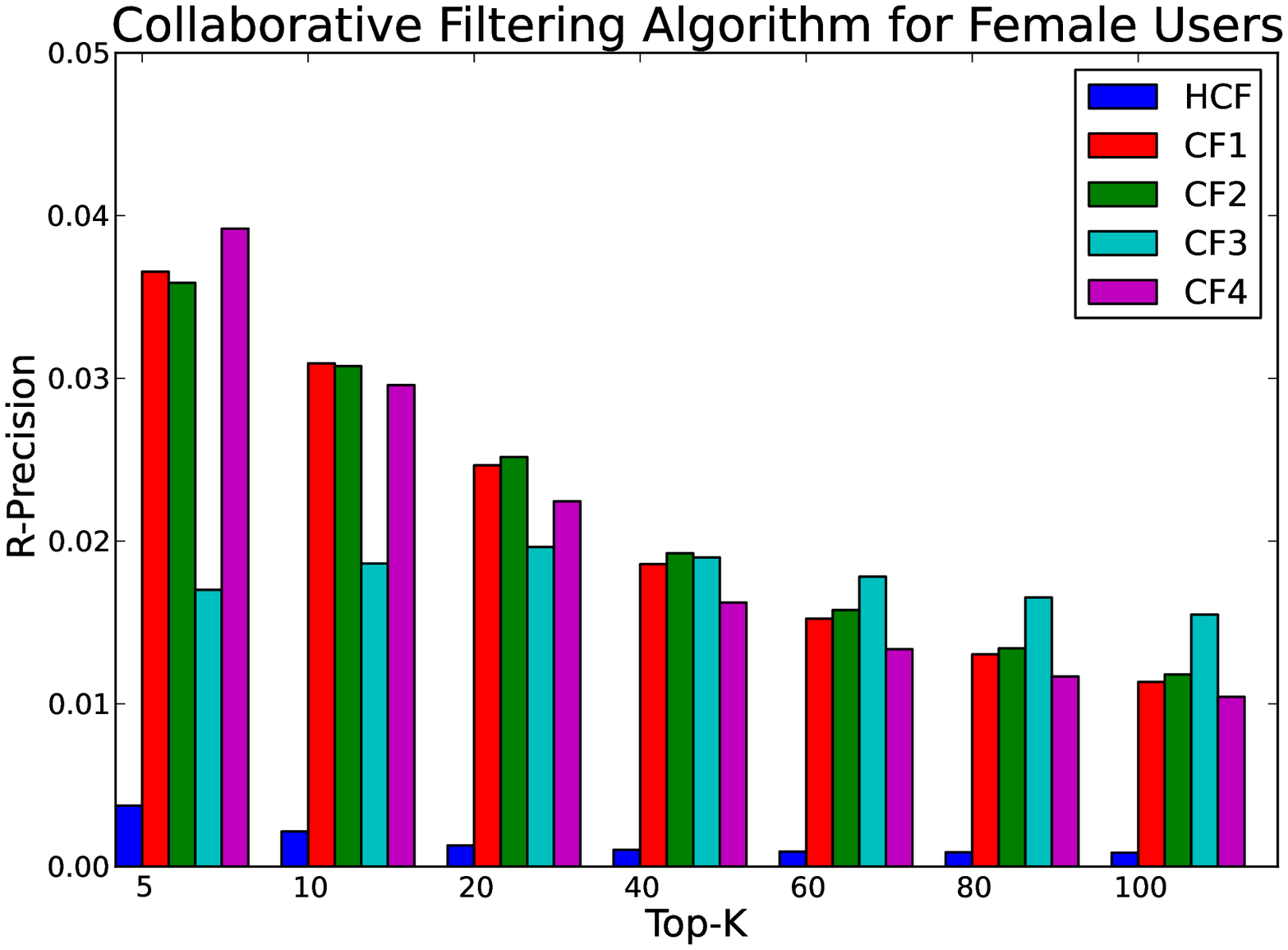}}\hfill
\subfloat[\label{fig:FCFRR}]
  {\includegraphics[width=.5\linewidth]{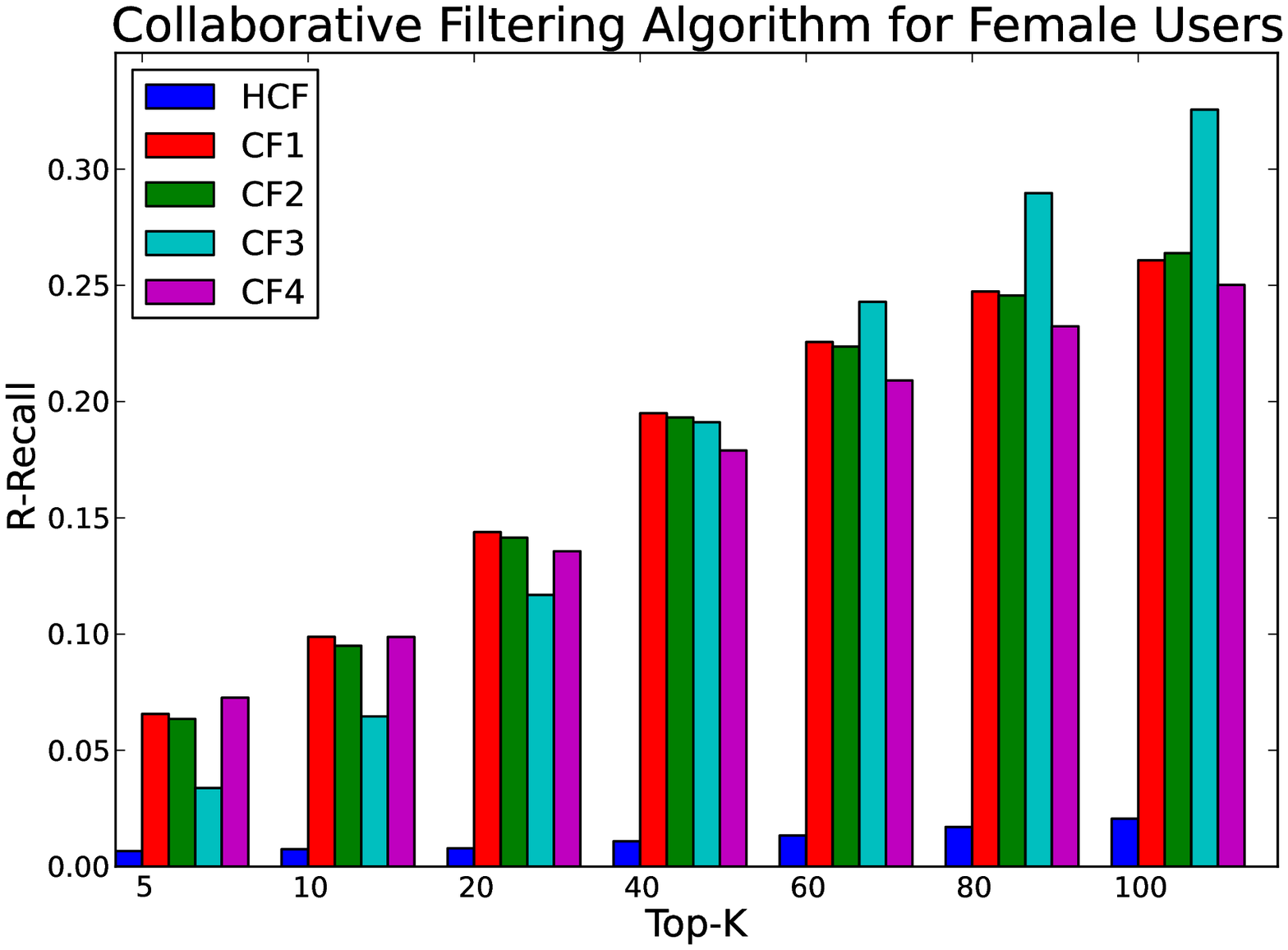}}\hfill
\caption{$R\text{-}Precision$ and $R\text{-}Recall$ of collaborative filtering algorithms for female users.}
\label{fig:cf_r_f}
\end{figure}

The performance of the collaborative filtering-based algorithms for male and female users are shown in Figures \ref{fig:cf_r_m} and \ref{fig:cf_r_f}, respectively. The algorithms proposed in our paper (CF1-CF4) still achieve much better results than HCF. For male users, while CF4 still outperforms the other algorithms, the difference is not as pronounced as for $I\text{-}Precision$ and $I\text{-}Recall$ measures. For female users, CF1 and CF2 show
very similar behavior. Recall that CF1/CF2 captures the mutual interest/attractiveness between the service user and recommended users. This shows that learning the mutual interest and mutual attractiveness between two users have similar effects for recommending potential dates for females. Unlike for male users, CF4 does not outperform other algorithms for female users. On the contrary, when the recommendation list (K) becomes large, CF3 starts to outperform the other algorithms. Recall that CF3 is symmetric to CF4, capturing the interest from recommended users in the service user and the attractive of the service user to the recommended users. The results indicate that when females look for potential dates, they are more conscientious to their own attractiveness to the other side of the line and the other sides' interest in them.

\subsection{Ranking Effectiveness}


\begin{figure}
\subfloat[\label{fig:MRANKN}]
  {\includegraphics[width=.5\linewidth]{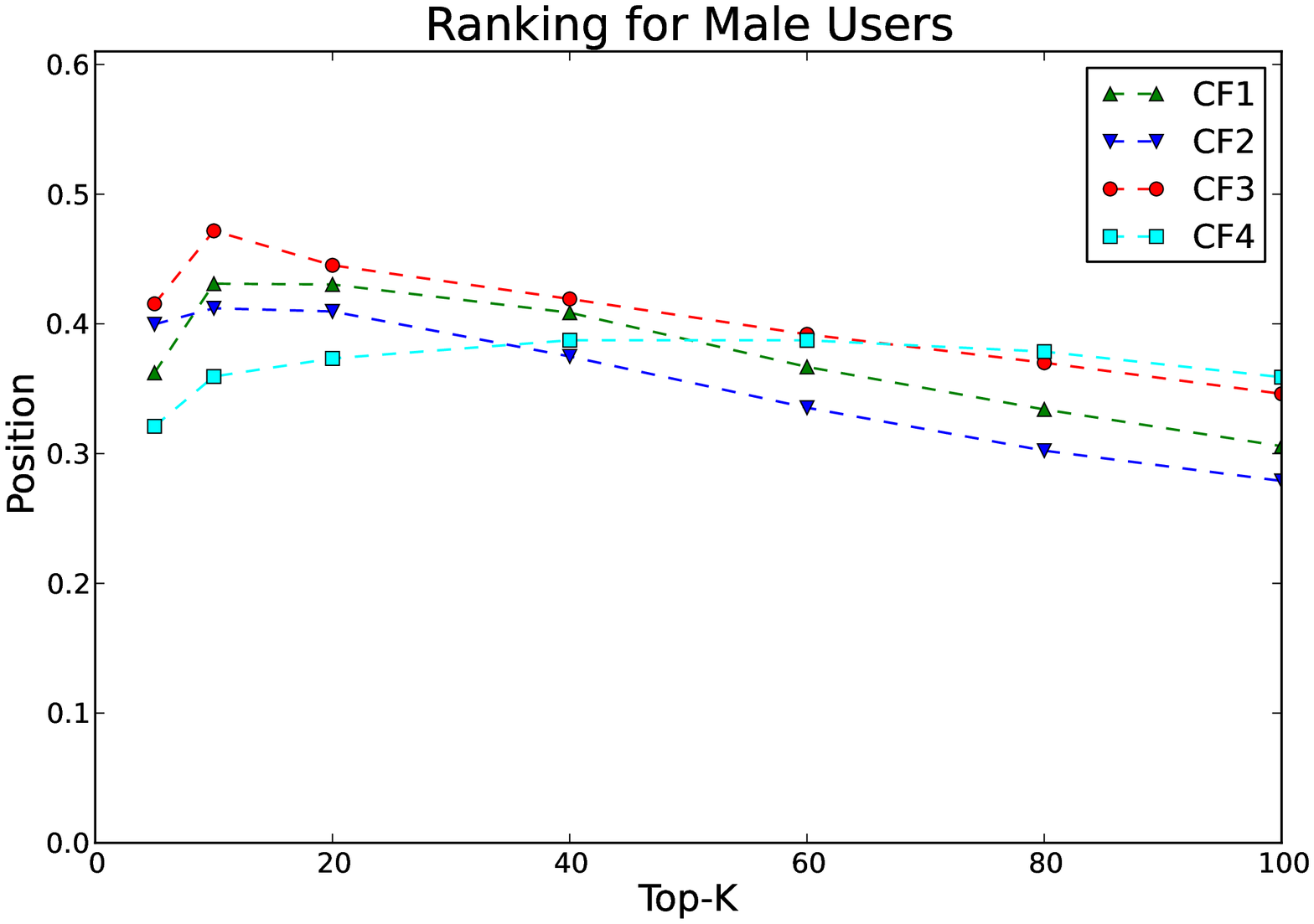}}\hfill
\subfloat[\label{fig:FRANKN}]
  {\includegraphics[width=.5\linewidth]{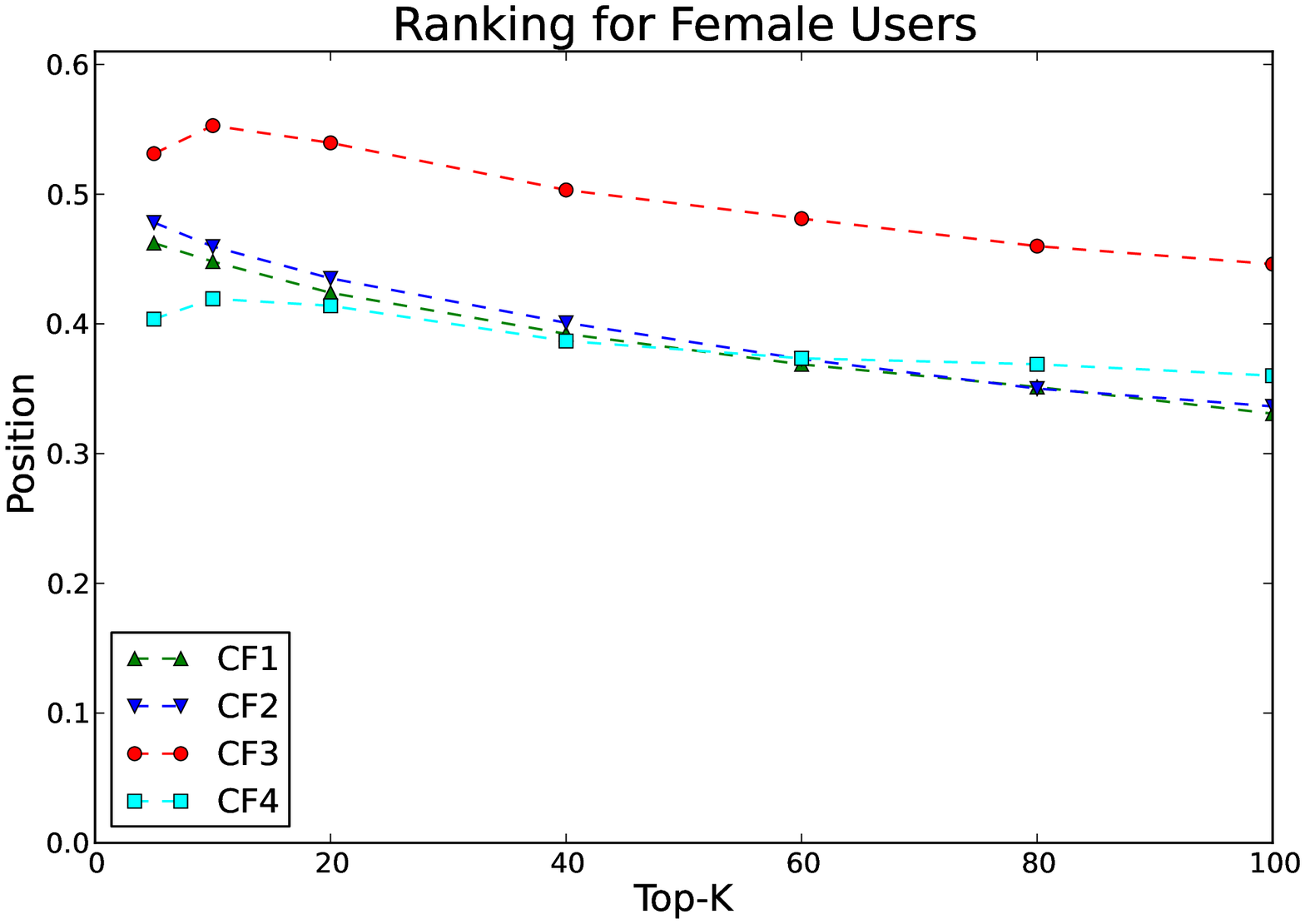}}\hfill
\caption{Average effective recommendation position of proposed recommendation algorithms for (a) male users and (b) female users.}
\label{fig:rank}
\end{figure}

In addition to precision and recall, the relative positions of relevant recommendations are also an important measure for a recommendation system \cite{Hannon10Recommending}. Relevant recommendations are defined as users in the recommendation list who have actually exchanged messages with the service user, i.e., the service user has followed the recommendation by contacting the recommended user who in turn has replied to the service user. Since a user usually looks at the recommendation list from top to bottom, a recommendation system ranking the relevant recommendations in top positions should be considered better than those with similar performance in precision and recall but ranking the relevant recommendations in lower positions.

\begin{figure}
\subfloat[\label{fig:age}]
  {\includegraphics[width=.5\linewidth]{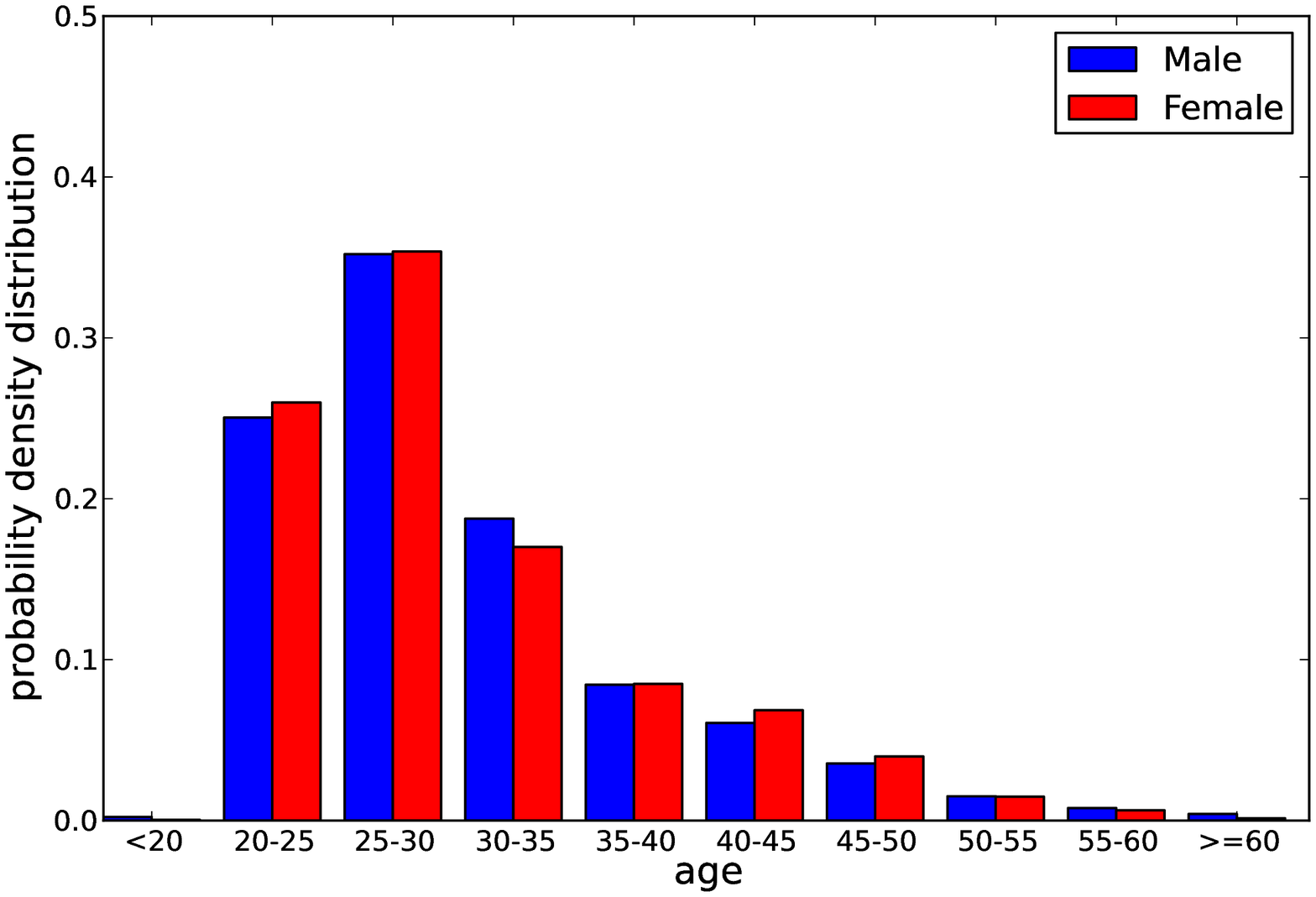}}\hfill
\subfloat[\label{fig:age_send}]
  {\includegraphics[width=.5\linewidth]{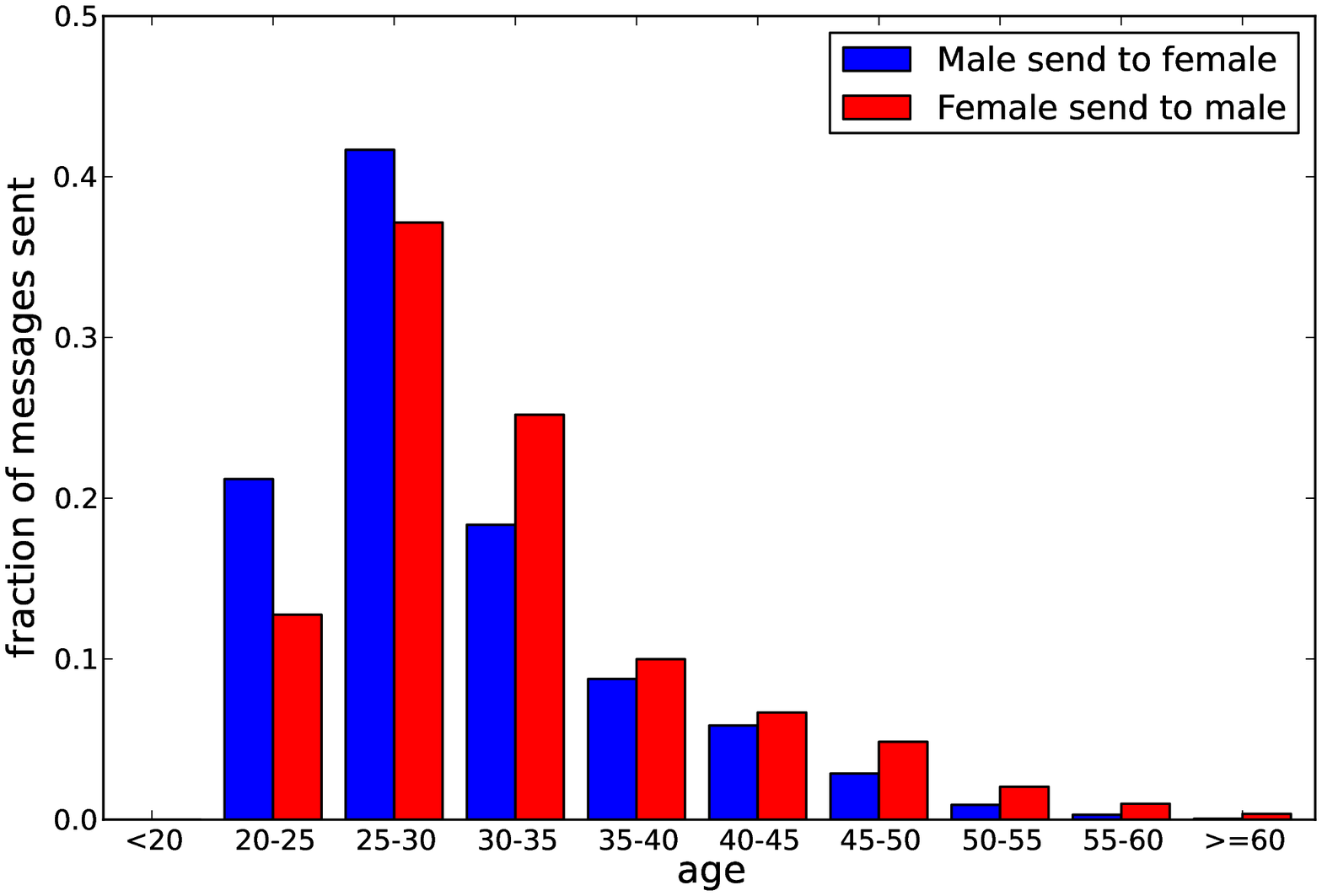}}\hfill
\caption{(a) Age distribution of all users, (b) Age distribution of messages sent.}
\label{fig:age_behavior}
\end{figure}

Figure \ref{fig:rank} plots the average positions of the relevant recommendations in the recommendation list (normalized by the size of recommendation list).  All of these algorithms rank the effectively recommended users in the top 30\% to 50\% of the recommendation list except for CF3 for female users which ranks the relevant recommendations around the halfway of the recommendation list. Note that we only present the reciprocal ranking effectiveness of CF1-CF4 algorithms as the other algorithms (HCF, CB1 and CB2) perform much worse than these algorithms. 
 

\subsection{Discussions}

To illustrate the relatively poor performance of the content-based algorithms when compared with the collaborative filtering-based approaches, we examine the effectiveness of using a user's attributes to determine whether another user would send a message to him or her.

\begin{figure}
\subfloat[\label{fig:education}]
  {\includegraphics[width=.5\linewidth]{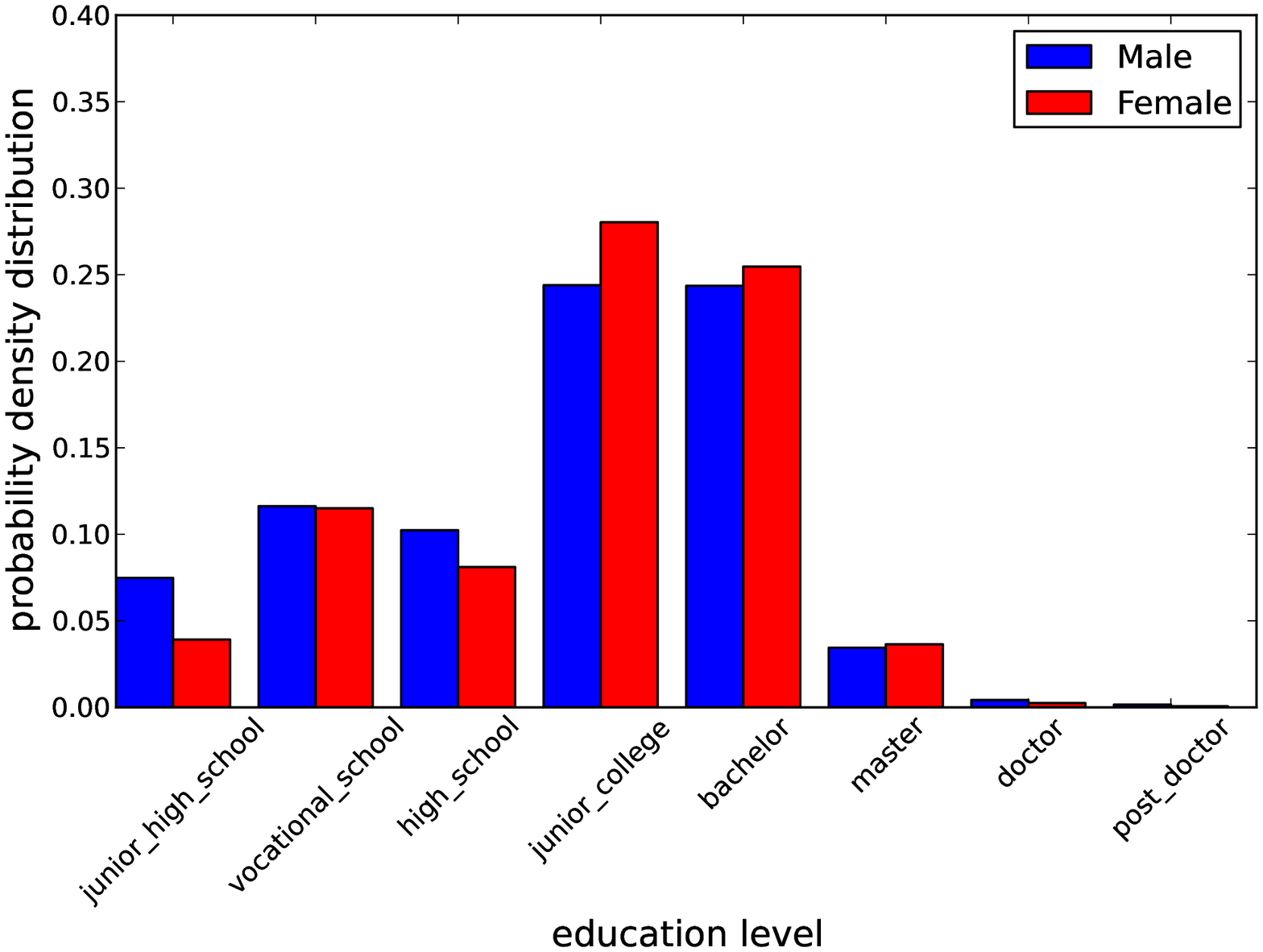}}\hfill
\subfloat[\label{fig:education_send}]
  {\includegraphics[width=.5\linewidth]{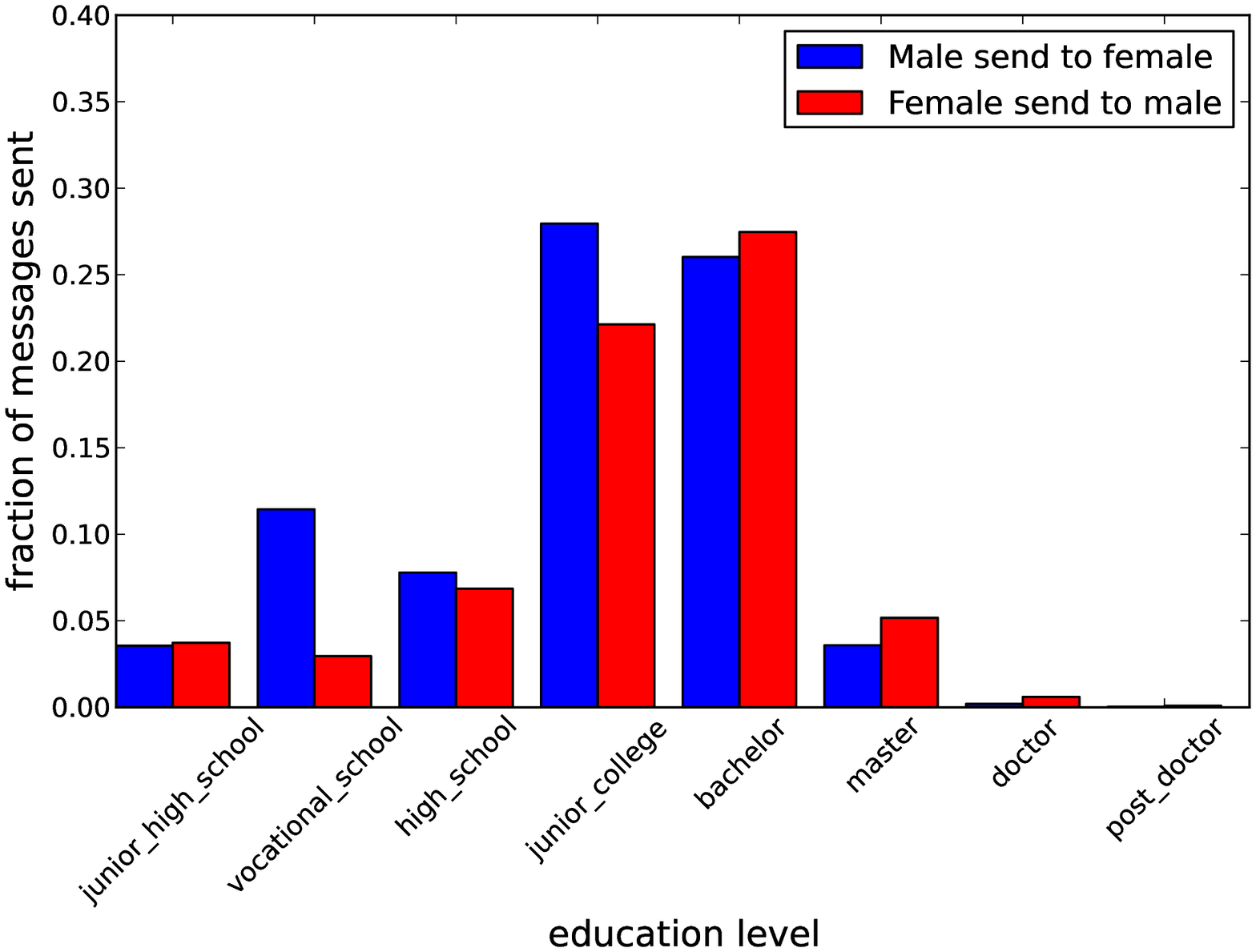}}\hfill
\caption{(a) Education distribution of all users, (b) Education Status distribution of messages sent.}
\label{fig:education_behavior}
\end{figure}

\begin{figure}
\subfloat[\label{fig:marriage}]
  {\includegraphics[width=.5\linewidth]{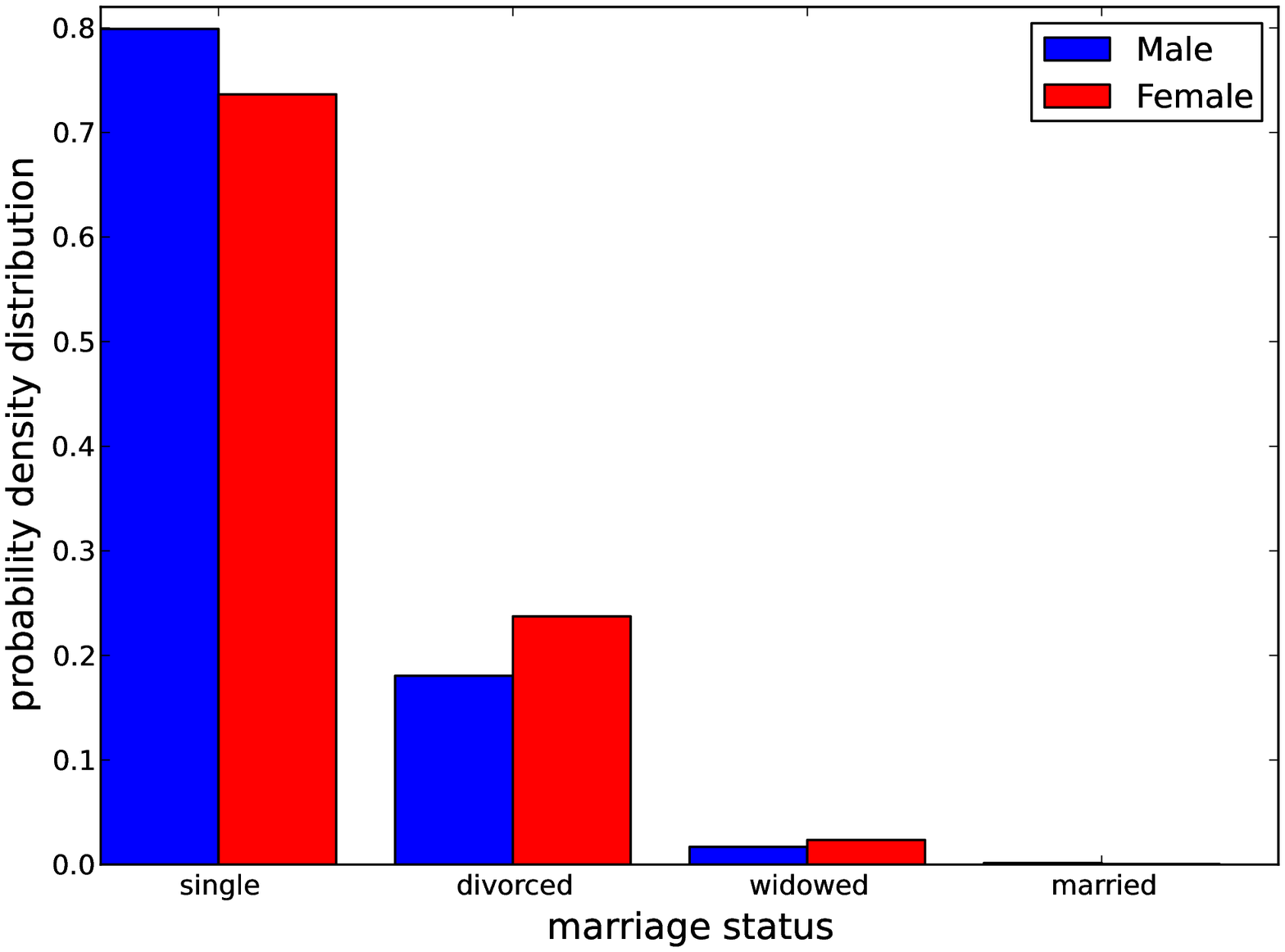}}\hfill
\subfloat[\label{fig:marriage_send}]
  {\includegraphics[width=.5\linewidth]{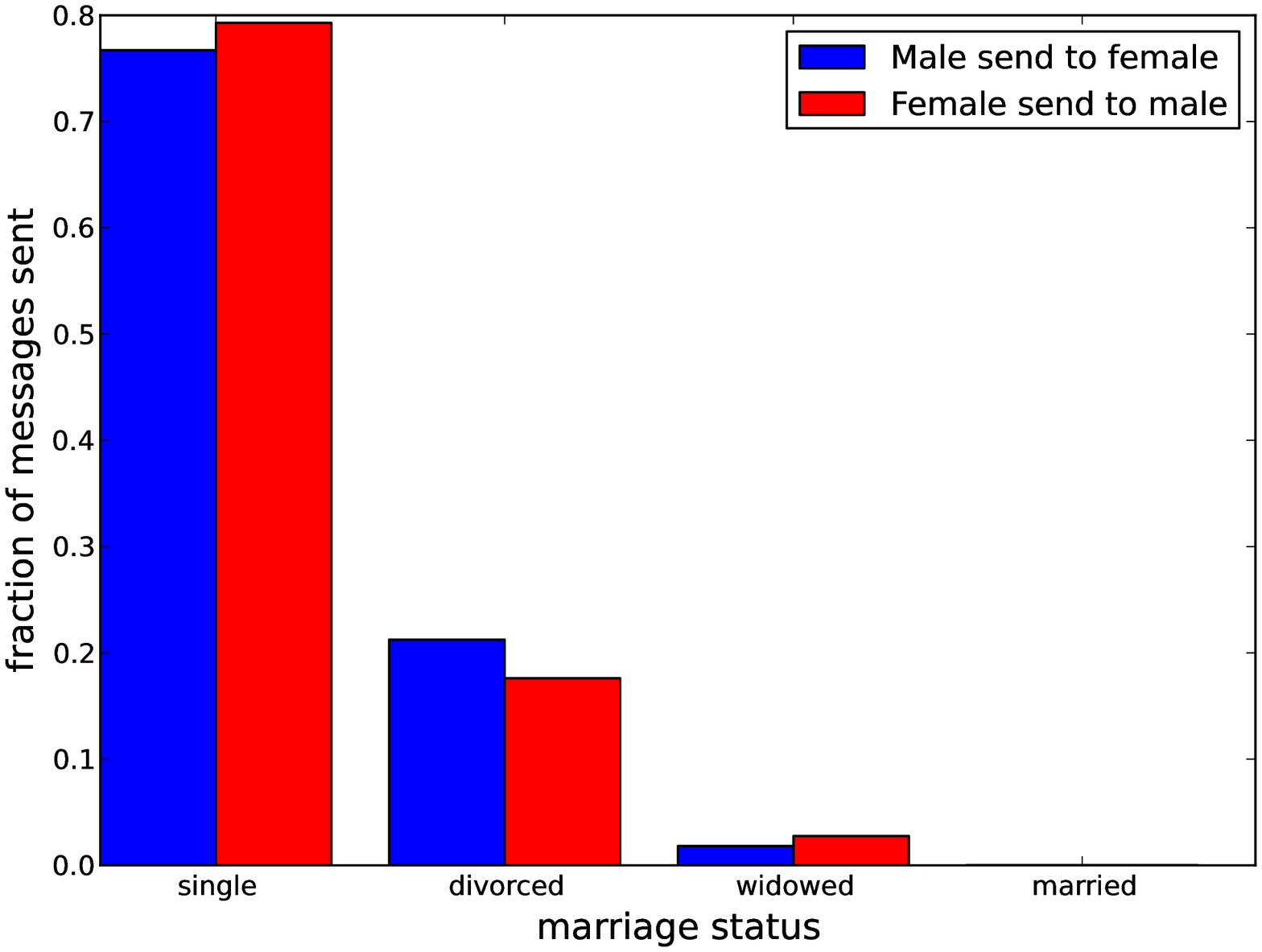}}\hfill
\caption{(a) Marriage distribution of all users, (b) Marriage Status distribution of messages sent.}
\label{fig:marriage_behavior}
\end{figure}

We plot several attribute distributions of all users as well as those of users who received messages for the age (Figure \ref{fig:age_behavior}), education level (Figure \ref{fig:education_behavior}), and marriage status (Figure \ref{fig:marriage_behavior}) attributes. We observe that for these attributes, the distributions for message receivers are quite similar to those for all users with small Bhattacharyya distances, indicating that these attributes are not very effective in making recommendations. Specifically, the Bhattacharyya distance between the two age distributions is 0.122 for males, and 0.064 for females. The Bhattacharyya distance between the two education distributions is 0.155 for males and 0.011 for females. For marriage status, the Bhattacharyya distance is blow 0.032 for both males and females. We also examined other attributes, which show small Bhattacharyya distances too.  For users in the age range of 20-30,  junior college or bachelor degrees, or single marriage status, it is difficult to distinguish them as these users constitute the majority of the population. 

\section{Conclusions} \label{sec:conclusion}
Matching users with mutual interest in each other is an important task for online dating sites. In this paper, we propose a set of similarity based reciprocal recommendation algorithms for online dating. We introduce several similarity messures that characterize the attractiveness and interest between two users, and select most compatible users for recommendations. We evaluate the performance of our proposed algorithms on a large dataset obtained from a major online dating site in China. Our results show that the collaborative filtering-based algorithms achieve much better performance than content-based algorithms in both precision and recall, and both significantly outperform previously proposed approaches. Our results also show that male and female users behave differently when it comes to looking for potential dates. In particular, males tend to be focused on their own interest and oblivious towards their attractiveness to potential dates, while females are more conscientious to their own attractiveness to the other side on the line.

\label{references}

\end{document}